\documentclass[review]{elsarticle}

\usepackage{lineno,hyperref}
\modulolinenumbers[5]

\journal{Astronomy and Computing}

\biboptions{authoryear}

\usepackage{graphicx}
\usepackage[printonlyused,nohyperlinks]{acronym}
\newcommand\MyBox[1]{%
    \fbox{\parbox[c][1cm][c]{3.5cm}{\centering #1}}%
}
\newcommand\MyVBox[1]{%
    \parbox[c][0.5cm][c]{0.5cm}{\centering\bfseries #1}%
}  
\newcommand\MyHBox[2][\dimexpr1.5cm+2\fboxsep\relax]{%
    \parbox[c][0.5cm][c]{#1}{\centering\bfseries #2}%
}  
\newcommand\MyTBox[4]{%
    \MyVBox{#1}
    \MyBox{#2}\hspace*{-\fboxrule}%
    \MyBox{#3}\par\vspace{-\fboxrule}%
}   
\usepackage{booktabs}
\usepackage{multirow}

\begin{document}

\begin{frontmatter}

\title{Automatic classification of eclipsing binary stars using deep learning methods}


\author[upjs]{Michal \v{C}okina} 

\author[tuke]{Viera Maslej-Kre\v{s}\v{n}\'akov\'a}

\author[tuke]{Peter Butka\corref{mycorrespondingauthor}}
\cortext[mycorrespondingauthor]{Corresponding author}
\ead{peter.butka@tuke.sk}

\author[upjs]{\v{S}tefan Parimucha} 

\address[upjs]{Department of Theoretical Physics and Astrophysics, Faculty of Science, Pavol Jozef \v{S}af\'arik University in Ko\v{s}ice, Park Angelinum 9, 040 01 Ko\v{s}ice, Slovakia}
\address[tuke]{Department of Cybernetics and Artificial Intelligence, Faculty of Electrical Engineering and Informatics, Technical University of Ko\v{s}ice, Letn\'a 9, 042~00 Ko\v{s}ice, Slovakia}

\begin{abstract}
In the last couple of decades, tremendous progress has been achieved in developing robotic telescopes and, as a result, sky surveys (both terrestrial and space) have become the source of a substantial amount of new observational data. These data contain a lot of information about binary stars, hidden in their light curves. With the huge amount of astronomical data gathered, it is not reasonable to expect all the data to be manually processed and analyzed. Therefore, in this paper, we focus on the automatic classification of eclipsing binary stars using deep learning methods. Our classifier provides a tool for the categorization of light curves of binary stars into two classes: detached and over-contact. We used the ELISa software to obtain synthetic data, which we then used for the  training of the classifier. For evaluation purposes, we collected 100 light curves of observed binary stars, in order to evaluate a number of classifiers. We evaluated semi-detached eclipsing binary stars as detached. The best-performing classifier combines bidirectional Long Short-Term Memory (LSTM) and a one-dimensional convolutional neural network, which achieved $98\%$ accuracy on the evaluation set. Omitting semi-detached eclipsing binary stars, we could obtain $100\%$ accuracy in classification. 
\end{abstract}


\end{frontmatter}

\section{Introduction}
\label{sec:introduction}
In recent years, photometric surveys have produced a large number of stellar light curves, which contain a large number of eclipsing binary stars. Eclipsing binary systems are part of a wide group of variable stars; their variability is caused by the mutual eclipses of two components during their orbital motion around a common center of gravity. An eclipse occurs when, from an observer's point of view,  one component is overlapped by the other. This phenomenon causes a decrease in the intensity of light emitted by the binary system. The resulting light curve reflects the processes in the system and their evolution with time \citep{prsa2018book}.

In our work, we consider light curves, which represent essential input data that can be obtained for the study of eclipsing binary stars. The light curves are displayed as graph which represents the brightness of an object over a certain period of time. The brightness might be shown in physical units (e.g., magnitude) or normalized to a certain value, thus unit-less. The $x$-axis which represents time progression is usually expressed in time units as Julian dates or more often are given in a dimensionless quantity called a photometric phase. When we study objects that have variable light intensity over time, such as eclipsing binaries, and/or supernovae, novae, or cataclysmic variable stars, light curves provide a valuable source of information and give us a straightforward way to be able to study such objects \citep{hurley}.

Binary stars -- especially eclipsing binaries -- are crucial objects in astrophysics. Binary star systems and multiple star systems represent more than half of all known stars in the Universe \citep{abt1983}. Analyses of their light-curves and other data, such as their radial velocities and spectra, allow us to determine the basic parameters of the components, such as stellar masses, temperatures, and ages, as well as the parameters relating to their orbits. The study of binary stars provides an opportunity to better understand the evolution of single stars and to explain many of the observed phenomena in the Universe. The most straightforward source of information about a binary system is its photometric light curve, where the first piece of information, which might be derived from such data, is a classification of the given source.

Eclipsing binary stars are types of extrinsic variable stars, where variations are due to external factors. To the observer, these stars may appear as a single point of light. Still, according to its variability in brightness and spectroscopic observations, this one point of light originates from two stars orbiting each other. Changes in the light intensity of eclipsing binary stars are caused by the fact that, from the perspective of the observer, one star passes in front of the other \citep{percy2007}. If we assume that the stars are spherical and have circular orbits, then we can easily estimate how the light changes with time and the eclipsing of the binary stars \citep{wilson1971realization, wilson1979eccentric, prsa02}.

There are two different ways to classify eclipsing binary stars. The first one divides light curves into three categories: The first type is EA type or Algol ($\beta$~Persei), the second is EB or $\beta$~Lyrae type, and the third is EW or W~UMa type. The given classification reflects the shapes of light curves and does not take into account of the physics hidden beyond the object of origin. The EA types of light curves are known for their plateaus in parts with maximal flux and two apparently significant flux minima. The EB light curve types show continual changes in measured radiation power. Differences in minima are usually a sign that the temperatures of both stars in the system are different. Light curves of type EW show a continual behavior, similar to EB types, but the minima are deeper than EB, and their absolute difference is small \citep{skleton2009}.

The second possibility for classification is based on the boundaries of the elaborated physical model using the components of the binary system itself. At present, the most commonly used model is the general Roche model. The generic approach has been elaborated by \cite{wilson1979eccentric}, which gives us four possible configurations. This classification scheme describes the amount of filling of Roche lobes \citep{kallrath01}:
\\

\begin{figure}[t]
    \centering
    \includegraphics[scale=.40]{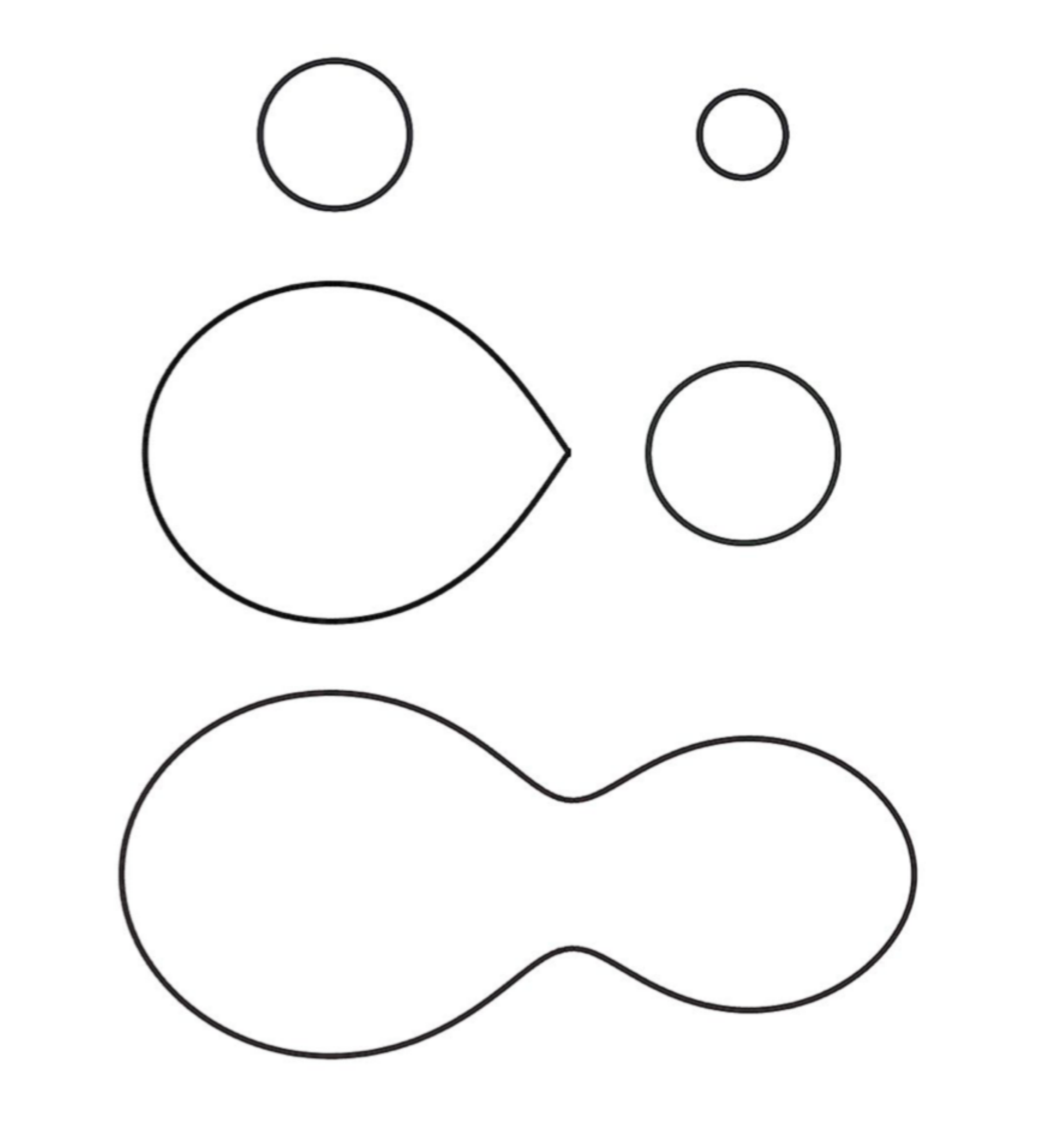}
    \caption{Examples of morphology of binary star systems based on the general Roche model. The figure represents detached, semi-detached, and over-contact binary systems, respectively, from top to bottom.}
    \label{fig:bs}
\end{figure}

\begin{figure}[!h]
    \centering
    \includegraphics[width=.48\textwidth]{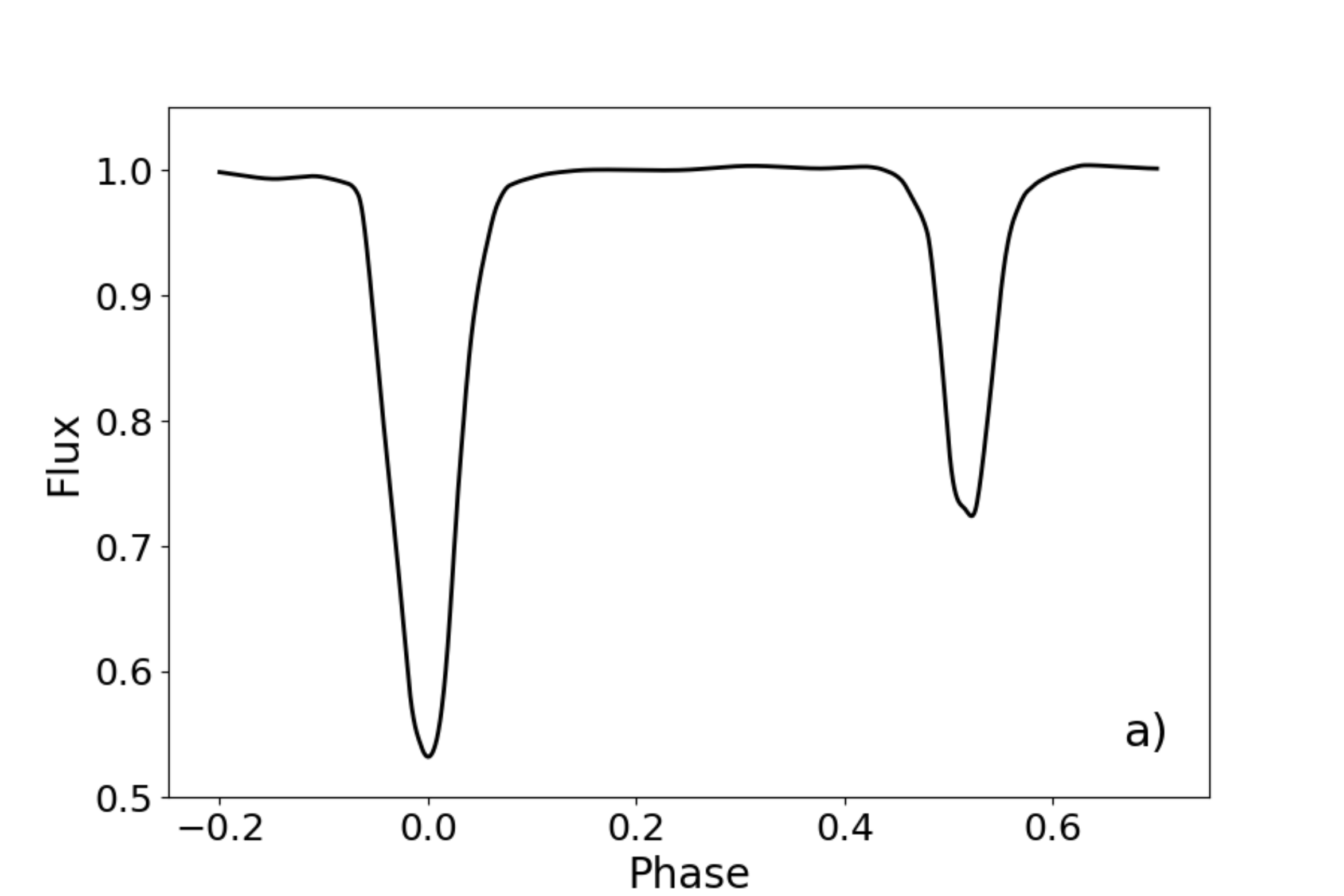}
    \includegraphics[width=.48\textwidth]{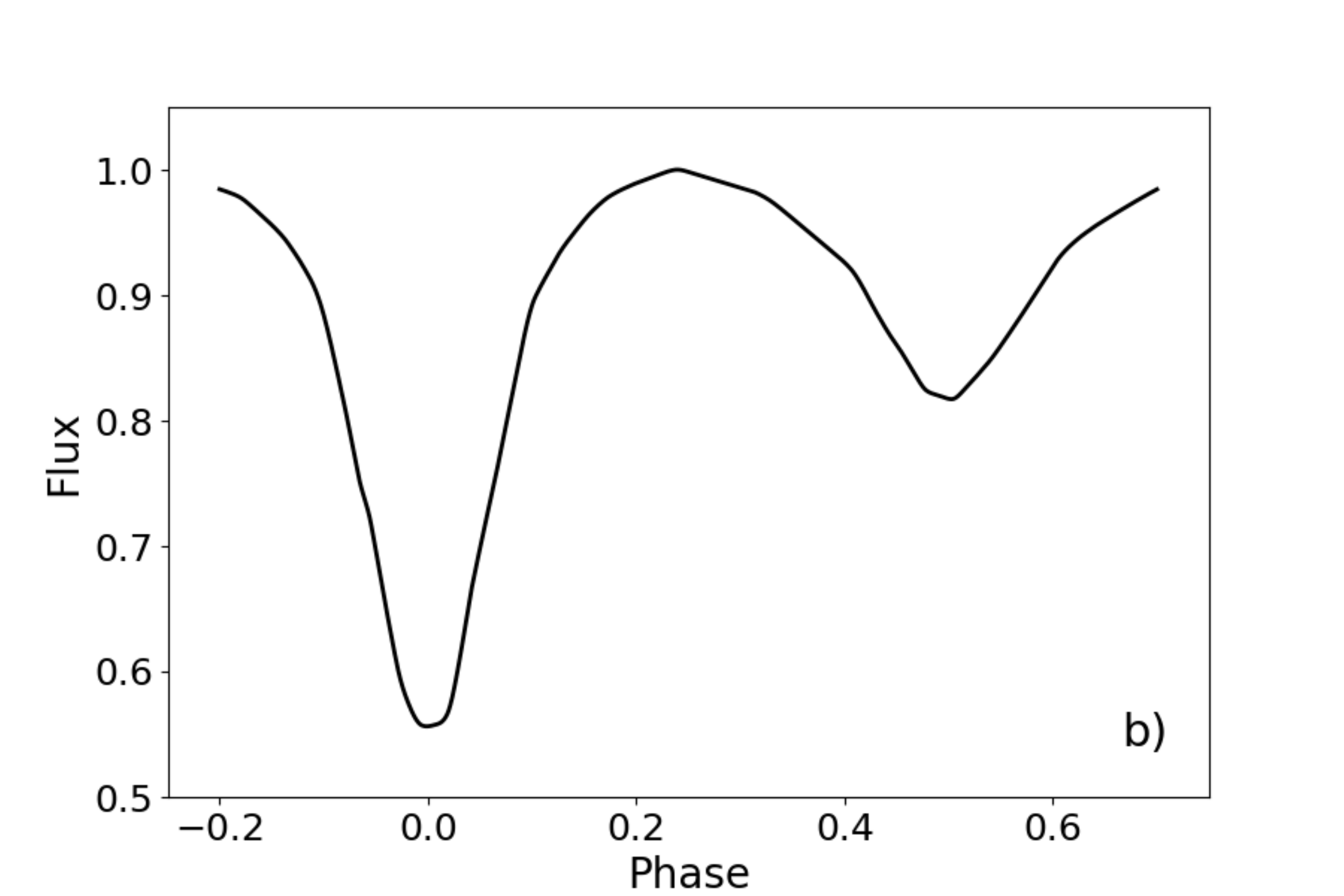}
    \qquad
    \includegraphics[width=.50\textwidth]{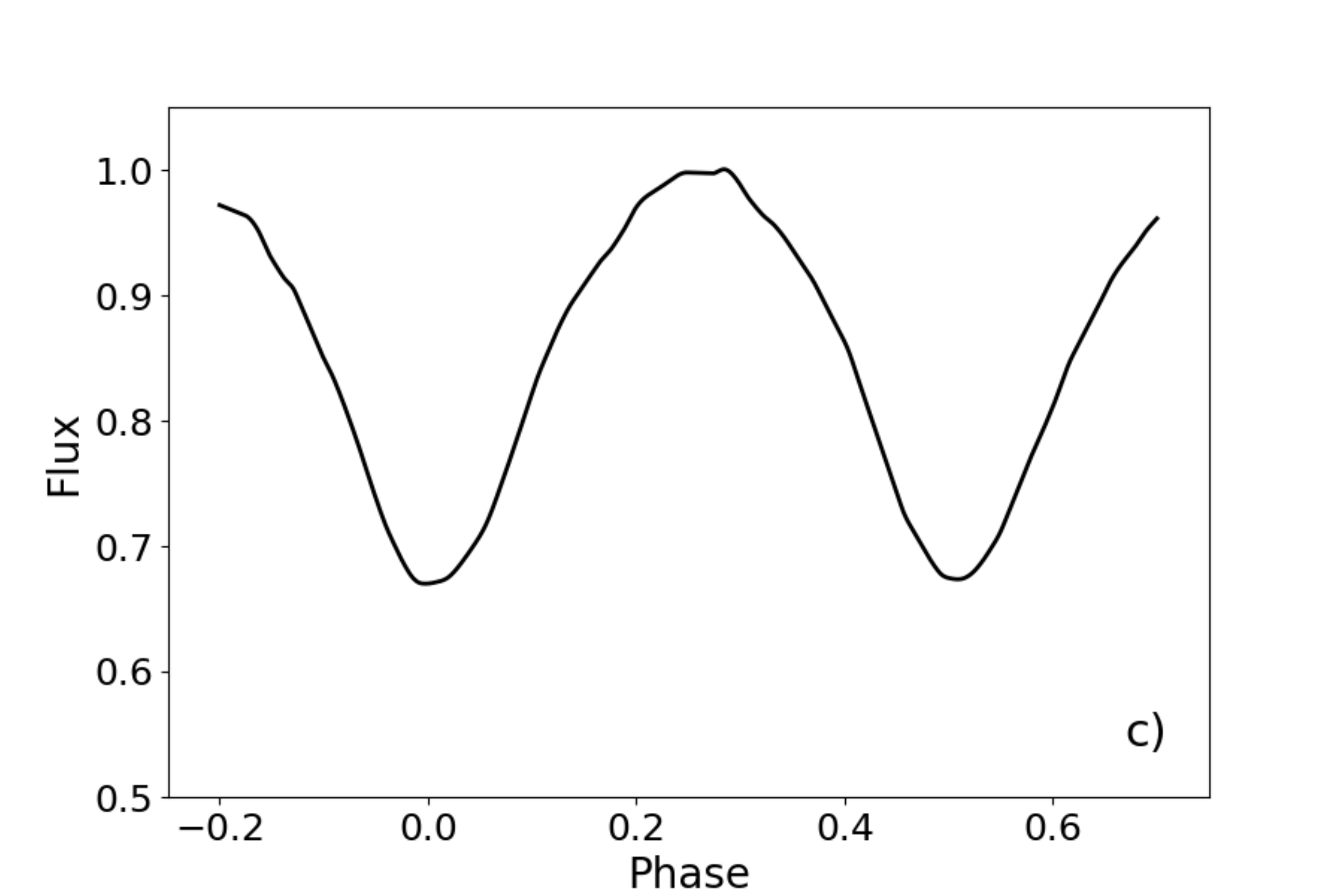}
    \caption{Representative light curves of \textbf{(a)} detached, \textbf{(b)} semi-detached and\textbf{(c)} over-contact binary stars.}
    \label{fig:morphology_examples}
\end{figure}

\begin{itemize}
\item{\textbf{Detached binary stars}}

Detached binary stars represent a pair of stars in which each star is located in its own Roche lobe (Figure~\ref{fig:bs}). The Roche lobe is the region around a star in a binary system within which orbiting material is gravitationally bound to that star. This region has the shape of a teardrop, the apex of which points towards the second star. In the Roche region, the gravitational force of the star is greater than that of the second one. The stars do not significantly affect each other and evolve independently. A representative example of a light curve is shown in Figure~\ref{fig:morphology_examples}a. More detailed discussion is provided by \citep{terrell2001} and \citep{malkov2007}.
\\
\item{\textbf{Semi-detached binary stars}}

Semi-detached binary stars are classified as where one of the stars fills its entire Roche lobe (the donor), while the other star (the recipient) does not. Gases from the surface of the donor Roche lobe are transferred to the second, rising star. Mass transmission dominates in this process. In many cases, the gas flowing to the receiving star forms a disk around it. In Figure \ref{fig:bs}, such a binary system is shown in the middle. A representative example of a light curve is shown in Figure~\ref{fig:morphology_examples}b.
\\
\item{\textbf{Over-contact binary stars}}

Over-contact binary stars are those in which both components forming the binary star system overflow their Roche lobes. The orbiting material in the Roche lobes of the two stars combines to create an envelope over the binary system as a whole. The friction of the orbiting material in the envelope inhibits orbital motion, which can lead to the stars eventually merging. In Figure \ref{fig:bs}, an over-contact system is shown at the bottom. A representative example of a light curve is shown in Figure~\ref{fig:morphology_examples}c.
\\
\item{\textbf{Double contact binary}}
There is also a fourth class, known as a double contact binary system, where each component fills its lobe exactly and at least one rotates super-synchronously. Such a state occurs for eccentric orbits or systems where the synchronicity parameter $F\neq1$. Thus double contact binary systems represent an extreme case in such a configuration. The physical origin comes from gas transfer and subsequent changes in the system \citep{wilson1994}.
\end{itemize}

\section{Motivation}

Nowadays, the problem of identification  of stellar parameters in binary star systems can be labeled as a big data problem. Recent development in observation techniques have led to the acquisition of huge amount of data sets that cannot be processed in real time. A lot of observed data might possess information about important phenomena happening in the Universe. It is crucial to be able to process data in real-time or at least as soon as possible. Human resources can not cover that, and we have to seek other, more sophisticated solutions.

Currently, researchers process and analyze a vast amount of data by hand with the help of developed tools, e.g.; Phoebe v1 \citep{prsa01}, Phoebe v2 \citep{prsa02}, ELC \citep{Orosz00}, etc. The determination process of parameters of close binary systems is not straightforward since it represents an inverse problem. The analysis is sensitive to the estimation of the initial values of parameters.
Such approach is a huge obstacle in automation, thus it is necessary to minimize or better to eliminate such human interaction, at least in initial analysis. The solution to the issue might be to involve machine learning algorithms in the process.

The ﬁrst step of the manual approach is to ﬁnd the morphological classiﬁcation since it determines the boundaries for surface potentials. This subsequently identifies  shapes of stars and all further related physical quantities, such as gravity acceleration, and temperature, etc. Usually, the appropriate implementation of developed tools relies on the chosen morphological classification and requires it for internal data processing. In this paper, we work on a suitable machine learning solution to determine morphological classification. We consider it to be the first and necessary step in the automation of light curve analysis.

\section{Related work}
The general classification of light curves is an important task in the study of different astronomical objects -- mostly related to the analysis of variable stars, but also to the analysis of objects in the Solar system or other planetary systems. In \cite{davis2018deep}, the authors provided a general classification of astronomical objects into five classes: asteroid, RR Lyrae (as an example of a pulsating variable star), Supernovae (as an example of a cataclysmic variable star), an eclipsing binary star, and a non-variable star. The classification was successful except for binary stars, which we address in the presented study.
In \cite{sarro2006automatic}, the authors presented a system that automatically classifies the light curves of eclipsing binary stars, where the classification aims to classify eclipsing binary stars according to their geometric configuration in a modified version of the classification scheme used. They used a Bayesian set of neural networks trained on data collected by the satellite Hipparcos for classification, where the data set consisted of seven different categories but also included the differentiation of eccentric binary systems and two types of morphologies of pulsating light curves. In practice, all classes in this scheme are based on the historical phenomenological classification of eclipsing binary curves into three groups (see Section~\ref{sec:introduction}): Algol, Beta Lyrae, and W Ursae Majoris. These systems are representatives of classes known as detached, semi-detached, and over-contact binary systems \citep{kallrath01}. The proposed model attempts to address the problems of the class heterogeneity and subjectivity of the traditional classification of light curves, which includes systems with different physical properties in the same group. Their classifier used etalons for groups, based on geometry in the sense that systems with the same geometric configuration are classified into the same group. The percentage of loss for their best architecture was $6.9 \pm 1.3\%$.

The morphological classification represents an initial approach for automatic light curve analysis. The more comprehensive goal is to determine the parameters of eclipsing binaries or at least find the initial parameters for common fitting approaches (e.g.; non-linear least square methods) or samplers (e.g.; Markov chain Monte Carlo). \citep{kochoska2020} attempted to tackle the problem with a nearest-neighbors search of most similar light curves in a pre-computed database. \citep{birky2020} also used the nearest-neighbors algorithm to classify the TESS data \citep{Ricker15}.  \citep{birky2020} showed that eclipsing binary types are distinguished from other periodic variables with a false-positive rate of 5 \% and a false-negative rate of 2\%.

\section{Data description}
\label{sec:datadesc}

\subsection{Synthetic data generator for light curves -- ELISa}
\label{sec:synthetic_data}
One of the problems in the area of  neural networks is the need of a large amount of representative input data. In a classification problem, the data usually consist of input features, usually vectors from space $R^n$ and related labels from space $R^m$. In some cases, if we know how to generate examples that simulate real-world objects, it is possible to solve this problem using synthetic data. As it is difficult to obtain a sufficient number of data set objects by annotating observational data of eclipsing binaries, we used the software ELISa \citep{elisa, cokina2021} (created by the co-authors of our paper), which is available online\footnote{https://github.com/mikecokina/elisa}.

The software applies known astrophysical knowledge about eclipsing binary stars. In general, it elaborates the theory provided in \cite{wilson1979eccentric} and improves it by using a geometric approach and discretization. The software tool was built to help users with star/binary modeling as well as with light curves and radial velocity curve analysis of the object. Its interface enables modeling capabilities to be accessed simply, so it is able to quickly generate synthetic light curves with the desired parameters and provides plenty of post-computed meta information such as the morphological classification of the modeled system. For the purpose of the learning procedure, we generated synthetic light curves in a phase range from -0.6 to 0.6. On-the-fly computed parameters, such as stellar radii, are restricted and the software does not take into account systems that violate the expected boundaries during the procedure. Boundaries for detached binaries were set based on the statistics published in \cite{eker_2014}. If any error occurred (e.g., gravity acceleration or the temperature of any surface element was outside the possible interpolation boundary given in \cite{castelli2004new}), such a combination of binary star parameters was rejected. The generator of the light curves was also restricted to keep surface potential the same for both components. Another restriction is that difference of the component's effective temperatures do not exceed 500 K. The full generator, with a description of the boundaries for parameters, is available on Github\footnote{https://github.com/mikecokina/elisa\_mlayer} and the entire set of parameters used for the synthetic generator is listed in Table \ref{table:gen_params_bounds}.

\begin{figure}[!h]
    \centering
    \includegraphics[width=.49\textwidth]{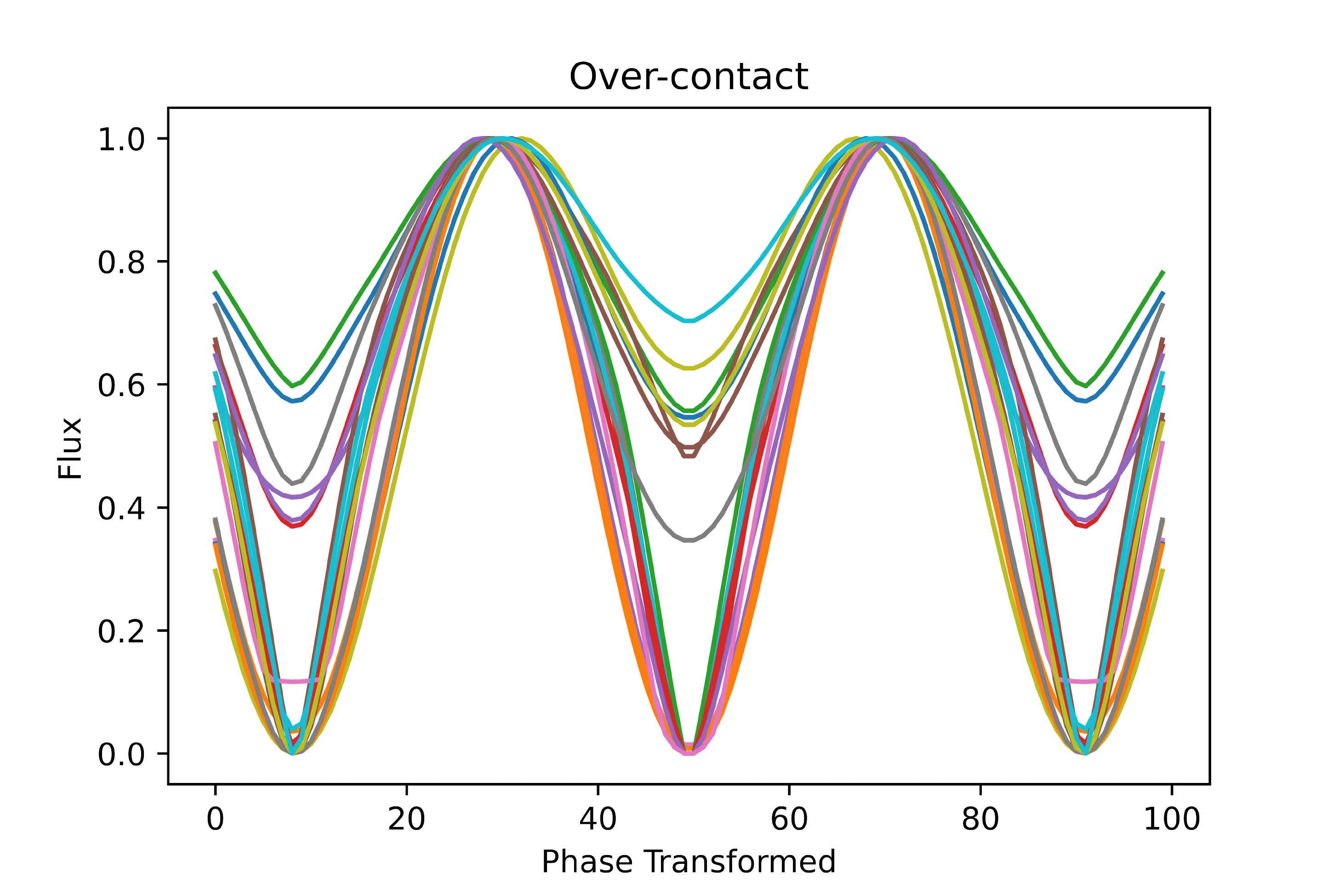}
    \includegraphics[width=.49\textwidth]{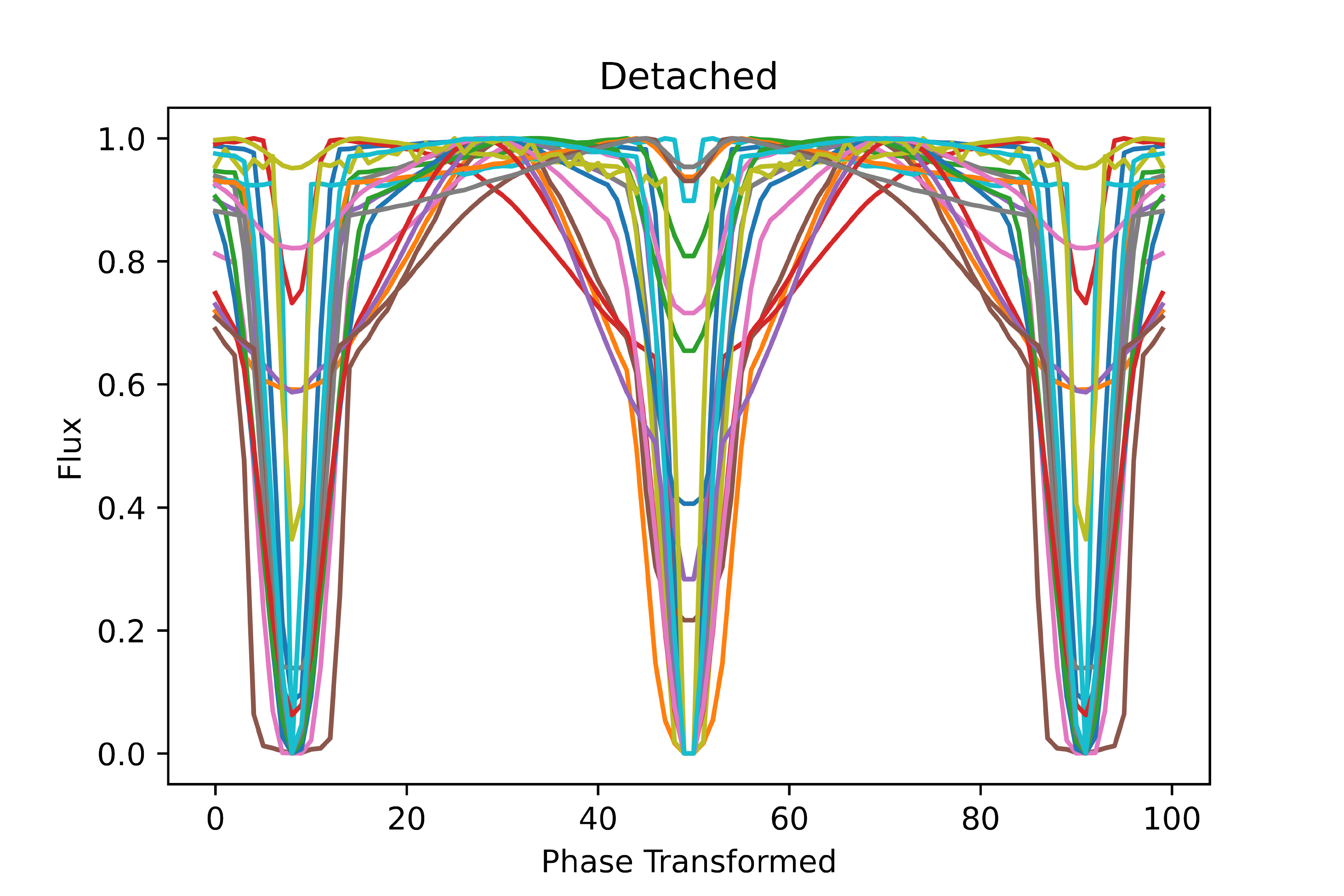}
    \caption{Demonstration of synthetic light curves of eclipsing binary stars generated by the ELISa software. The first graph shows over-contact binaries, while the second graph shows detached binaries.}
    \label{fig:s}
\end{figure}

We generated a total of 491,425 synthetic curves for neural network training (see Table \ref{t:pocet}). Figure \ref{fig:s} shows an example of these curves.

All generated light curves might be displayed as dependencies of normalized flux over orbital phase from phase -0.6 up to phase 0.6 with equidistant steps in phase. We used phase depended light curves to keep all the data constantly scaled in the $x$-axis. Flux in the $y$-axis was normalized for each data entry by its maximal value. It means that every processed light curve is normalized in flux and the maximal allowed value is equal to 1; then we have both axes dimensionless. In this case, input for the neural network does not require information about the $x$-axis and it might be omitted. We use as an input of the neural network a vector from space $R^{100}$. We consider equidistant steps between values of the orbital phase each related to the flux value in the vector. In other words, there exists a unique mapping from linear space [-0.6, 0.6] to our feature space [1, 100] and we do not have to take into account the $x$-axis and work only with normalized flux arranged in the feature vector.

\begin{table*}
\caption{Parameters and their boundaries used for the light curve generator, where $M$ stands for mass, $P$ for period, $i$ for orbital inclination, $\Omega$ for surface potential value, $T$ for effective temperature, $F$ for synchronicity factor, $A$ for albedo, $\beta$ for gravity darkening coefficient, and $[M/H]$ stands for metallicity.}
\label{table:gen_params_bounds}
    \centering
        \begin{tabular}{c c c c}
        \hline
        Parameter & Boundaries & Step & Unit\\   
        \hline\hline
        detached & & & \\
        \hline
        $M_i$, $i\in {1,2}$ & [0.2, 3.5] & 0.5 & $M_\odot$ \\
        P & [1, 40] & 5 & d \\
        i & [70, 90] & 5 & deg \\
        $\Omega_i$, $i\in {1,2}$ & [2, 7] & 1 & - \\
        $T_i$, $i\in {1,2}$ & [4000, 15000] & 2000 & K \\
        \hline
        over-contact & & & \\
        \hline
        $M_i$, $i\in {1,2}$ & [0.20, 3.0] & 0.25 & $M_\odot$ \\
        P & [0.2, 0.9] & 0.1 & d \\
        i & [30, 90] & 5 & deg \\
        $\Omega_1 = \Omega_2$ & [2.00, 3.75] & 0.25 & - \\
        $T_i$, $i\in {1,2}$ & [4000, 6500] & 500 & K \\
        \hline
        e & [0, 0] & - & - \\
        $F_i$, $i\in {1,2}$ & [1, 1] & - & - \\
        $A_i$, $i\in {1,2}$ & 0.5 if $T_i$ \textless 6500 else 1.0 & - & - \\
        $\beta_i$, $i \in {1,2}$ & 0.25 if $T_i$ \textgreater 6500 else 0.09 & - & - \\
        $[M/H]_i$, $i \in {1,2}$ & [0, 0] & - & - \\
        \hline
        \end{tabular}
\end{table*}

\begin{table}[!h]
\centering
\caption{Binary type occurrences in training and evaluation data sets. The training set includes synthetic data generated by the ELISa software. The evaluation set includes light curves of observed binaries.}
\begin{tabular}{lcc}
\toprule
\textbf{Type}	& \textbf{Training set}& \textbf{Evaluation set}	 \\
\midrule
Detached        & 445,556  & 53\\
Over-contated   & 45,869  & 47 \\ \midrule
Total & 491,425 & 100\\ 
\bottomrule
\end{tabular}
\label{t:pocet}
\end{table}

\subsection{Observational data of eclipsing binary stars}

We collected 100 light curves from observations to test our classifier. Since our testing sets contain observational data instead of synthetic ones as in the training set, we will use a term evaluation set for them throughout the paper. Therefore, our evaluation set contained the light curves of 39 detached, 14 semi-detached binaries, and 47 over-contact eclipsing binaries. As semi-detached binaries, in principle, represent detached binaries, we evaluated semi-detached observation light curves as detached ones. The morphological classifications of the chosen objects were determined based on available published catalogues \citep{southworth2015, malkov2006, pribulla2003}. We obtained data from related papers from the catalogue if available online. The data we found are usually depicted in different forms (time dependent, phase dependent, in normalized flux, in magnitude, etc.) but all the necessary information was provided alongside light curves, so we normalized them to match the synthetic data in the manner described in Section \ref{sec:synthetic_data}. We were not able to find data points for several light curves, thus we performed a digitization of available charts presented in published papers and as in the previous case, we applied the normalization procedure.

\begin{figure}[!h]
    \centering
    \includegraphics[width=.49\textwidth]{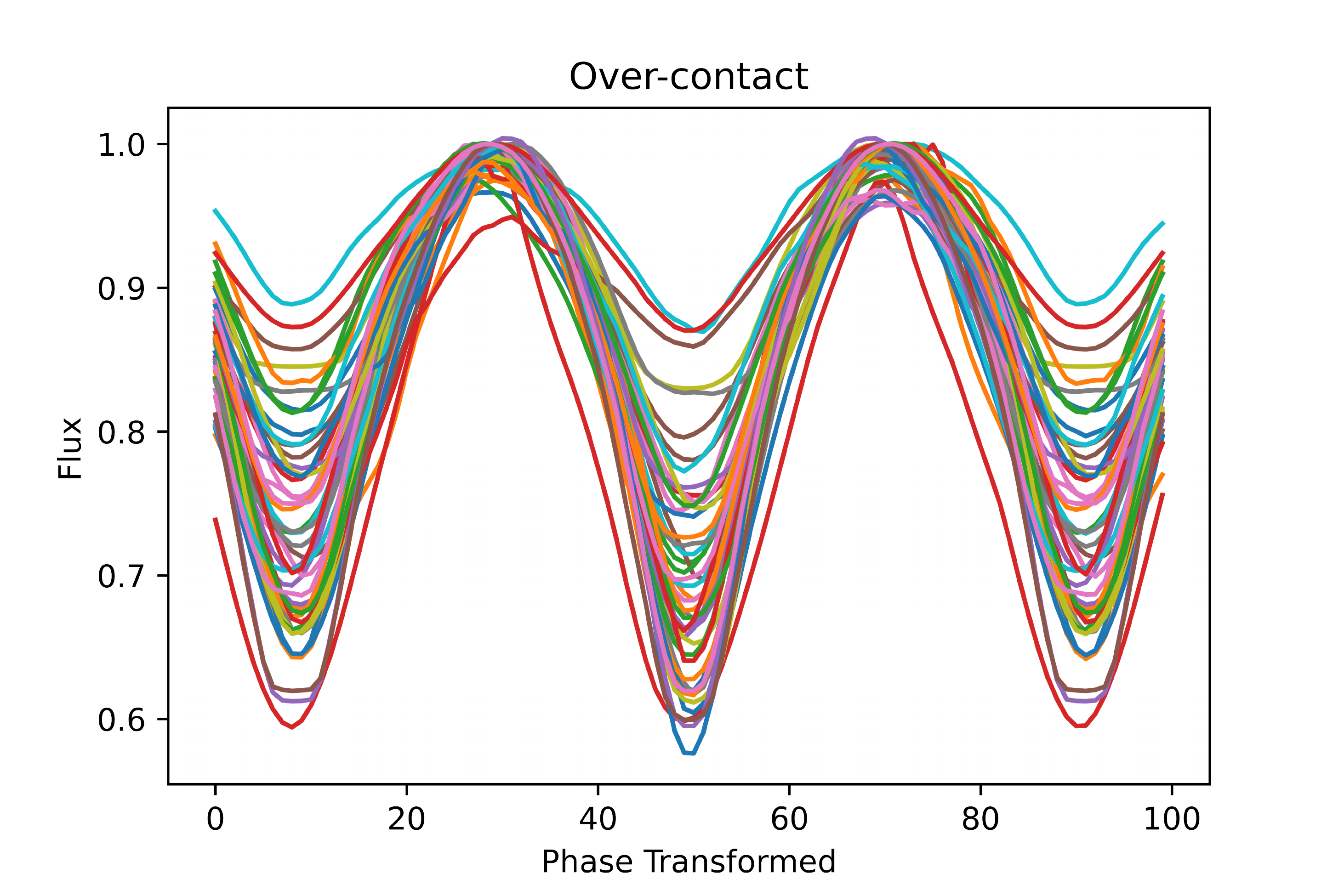}
    \includegraphics[width=.49\textwidth]{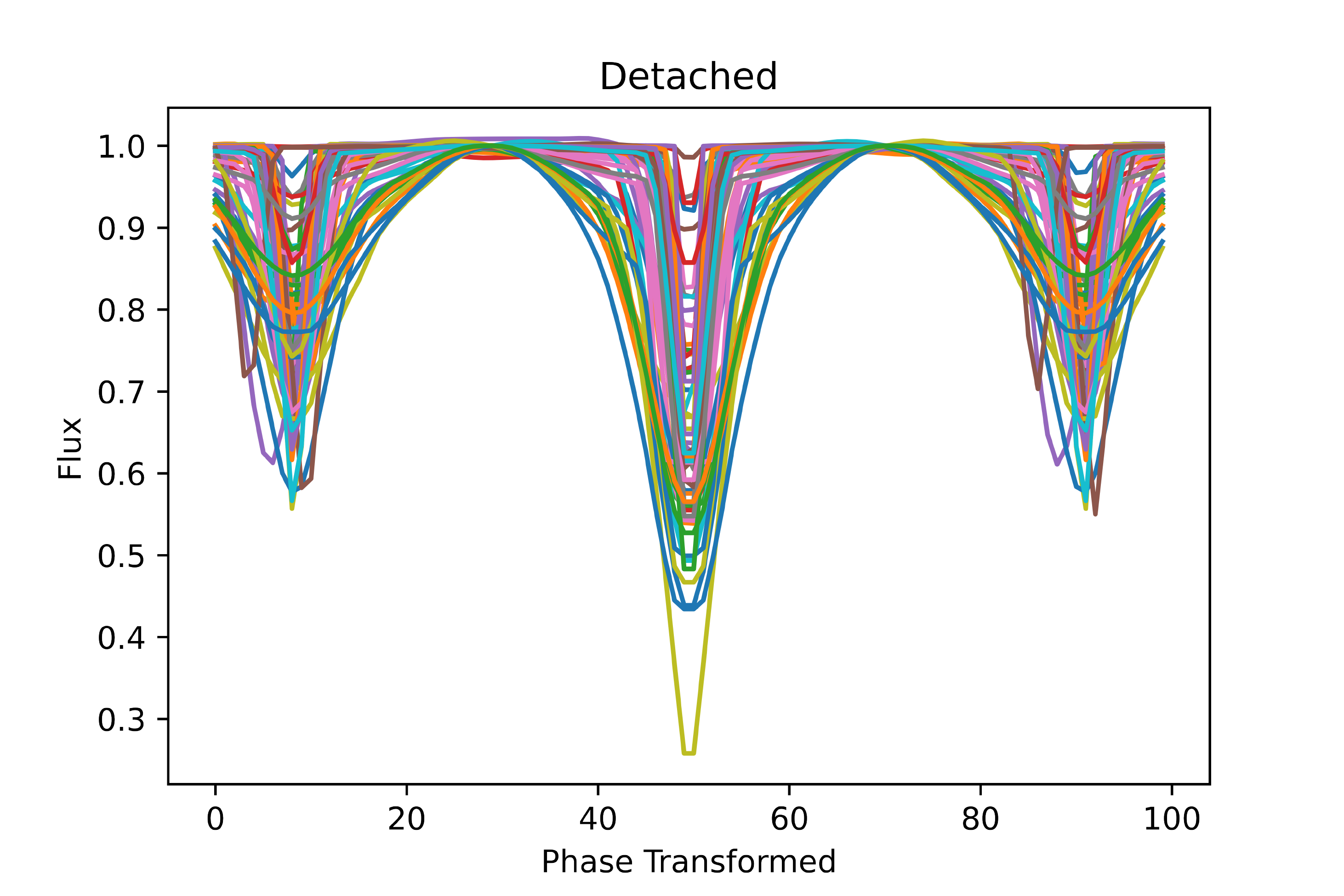}
    \caption{Demonstration of observational light curves of eclipsing binary stars. The first graph shows over-contact binaries, while the second graph shows detached binaries (including semi-detached eclipsing binary stars).}
    \label{fig:obs}
\end{figure}

Table \ref{tab:obsetvation} lists the names of the eclipsing binary stars whose light curves we used to evaluate the classifier. We show examples of such light curves in two graphs. The first graph in Figure \ref{fig:obs} shows over-contact binaries, while the second graph shows detached and semi-detached observation light curves (the latter were evaluated as detached).

\begin{table}[!h]
        \caption{Morphological classification of objects used for evaluation of neural networks.}
        \label{tab:obsetvation}
    \centering
    \footnotesize{
     \begin{tabular}{llllll} 
    \toprule
       \multicolumn{6}{l}{\textbf{Detached -- evaluated as Detached}}\\
    \midrule
        AD And              & AD Boo             & AQ Ser   & AR Aur  
                            & AS Cam             & ASAS J045304-0700.4 \\%
        BK Peg              &  CD Tau             & CF Tau &
        CO And              & CV Boo             & CoRoT 102918586\\ %
        HD 71636            &  HS Aur             & HY Vir &
        IM Vir              & MY Cyg             & NY Cep  \\
        PT Vel              & PV Cas            & RT And &
        RT CrB              &  UZ Dra             & V1229 Tau \\
        V364 Lac            & V396 Cas             & V442 Cyg &
        V459 Cas            & V624 Her             & V785 Cep \\
        V885 Cyg            & VV Crv             & VV Mon &
        VZ Cep              &  WX Cep            & WZ Oph  \\
        Z Her               & ZZ Boo             &     $\beta$ Aur  \\
            \bottomrule
    \toprule
        \multicolumn{6}{l}{\textbf{Semi-detached -- evaluated as Detached}} \\
    \midrule
        AS Eri              & AT Peg             &  AX Dra &
        DD Mon              & DL Cyg            & TX Cet\\
        V Crt               & V1241 Tau         & V504 Cyg &
        XZ Cep              & XZ Cmi            & Z Vul \\
        ZZ Aur              & $\mu^1$ Sco       &  \\
    \bottomrule
    \toprule
       \multicolumn{6}{l}{\textbf{Over-contact -- evaluated as Over-contact} }\\
    \midrule
        AA UMa              &  AB And         &   AD Phe &
        AH Aur              &  AO Cam         &    AO Cas  \\
        AP Aur              & AQ Tuc         &  AU Ser  &
        AW UMa              & BH Cas         & BI CVn \\
        BL And              &  BO CVn          & BV Eri &
        CC Com              & CE Leo         &  CN And  \\
        CT Tau              & DK Cyg        &   EK Com &
        EQ Tau              &  FT Lup       &  QX And \\
        RS Col              & RT LMi         &  RW Com   &
        RZ Com              & SS Ari         &  SW Lac  \\
        SY Hor              & TU Boo          & TW Cet   &
        TY Boo              &  UZ Leo         & V1073 Cyg \\
        V535 Ara            & V676 Cen        & V677 Cen &
        V752 Cen            & V728 Her          & V729 Cyg  \\
        V861 Her            & V857 Her         &   WZ Cyg &
        XY Boo              & YY CrB           & \\

                \bottomrule
    \end{tabular}}

\end{table}

\section{Methodology}
\label{sec:methodology}
\subsection{Selected deep learning methods}
Artificial intelligence is one of the fastest expanding areas in computer science. Scientists and enthusiasts in this field have been continually working to improve it and, with new features, expand the possibilities of its application. Deep learning methods have achieved significant progress in many areas, such as computer vision and time-series processing. One of the most successful models in the area of deep learning is known as a convolutional neural network, which can process sequences with its one-dimensional variant and, therefore, it is suitable for the analysis of time-series data \citep{cnn}. Other models that can solve the problem of sequence learning are recurrent neural networks \citep{Goodfellow-et-al-2016} where, in addition to spatial structures, time structures also appear (i.e.,  several outputs can be given for one input, depending on the time sequence of the given input). The simplest solution architecturally is a neural network with a time shift, which can be trained with the standard backpropagation method \citep{rumelhart1995backpropagation}.

In the following subsections, we briefly describe neural networks, from the simplest feed-forward neural network through to convolutional neural networks, recurrent neural networks (including LSTM and GRU), and their bidirectional variants.

\subsubsection{Feed-forward neural network}

The most straightforward neural network, which forms the basis for more complex feed-forward neural networks, is called a perceptron. The perceptron receives input signals $\bar{x} = (x_{1}, x_{2}, . . . , x_{n+1})$ through synaptic weights constituting the weight vector $\bar{w} = (w_{1}, w_{2}, . . . , w_{n+1})$.

The input vector $x$ is called a pattern or shape. The output of the perceptron is given as a scalar (component) product of the weight and input vector, transformed by an activation function $f$. The constant  $b$ is the \emph{bias}, which is added to the activation function (thus affecting it) and is independent of the input parameters \citep{Goodfellow-et-al-2016}. The output of the perceptron $O$ is given by:

\begin{equation}\label{r:3}
O = f(\bar{w} \cdot \bar{x}) = f( \sum_{i = 1}^{n + 1}w_i x_i) + b.
\end{equation}

Due to the backpropagation method, multi-layer feed-forward neural networks came to the forefront of deep learning research. Most notably, the backpropagation method can be used to solve non-linear problems \citep{rumelhart1995backpropagation}. In the training phase, the algorithm updates the values of weights and the bias of neurons in the neural network based on the output loss.

\subsubsection{Convolutional Neural Network}

A Convolutional Neural Network (CNN) \citep{cnn} is a deep learning algorithm that assigns importance to various properties, objects, or patterns in the data of an input image or sequence and, thus, can distinguish them from each other. The training algorithm assigns significance to individual patterns or objects through the use of weights and biases. Mathematically, it uses a convolution operation, in which we use various filters (also called kernels) to extract specific characteristics and features. In the lower layers, for example, these may be the edges of objects in images while, in higher layers, these are more complex shapes. The mathematical formulation of this operation is:
     \begin{equation}
     S(i,j)=(K*I)(i,j)=\sum_m \sum_n I(m,n)K(i-m,j-n),
    \end{equation}
where $K$ is the kernel (i.e., filter) and $I$ is the input shape \citep{Goodfellow-et-al-2016}.

Another specific layer in a CNN is the pooling layer \citep{maxp}, which has the aim of reducing the number of parameters. The output of the pooling layer allows for under-sampling and simplifies the input to lower dimensions. There are several types of pooling layers, based on the operation applied to neighboring neurons, such as max pooling, average pooling, global max, or average pooling. In practice, a CNN combines convolution and pooling layers which, at the end of the convolutional part, provide a compact representation of the input in a lower dimension. Then, we can apply the decision part of the network; for example, by using a fully connected multi-layer feed-forward network or another algorithm for classification.   
In the case of time-series data, we can easily apply a CNN to these one-dimensional data sequences, such as in the case of acceleration and gyroscopic data for human activity recognition, analysis of financial time series, for the processing of textual data in a sequence of words, signal processing for seismic sensors, or to classify other sequences. In this case, we are talking about a one-dimensional convolutional neural network (1D CNN), in contrast to images (where the input is a two-dimensional matrix of pixels).

\subsubsection{Long Short-Term Memory}
\begin{figure}[ht!]
    \centering
    \includegraphics[width=\textwidth]{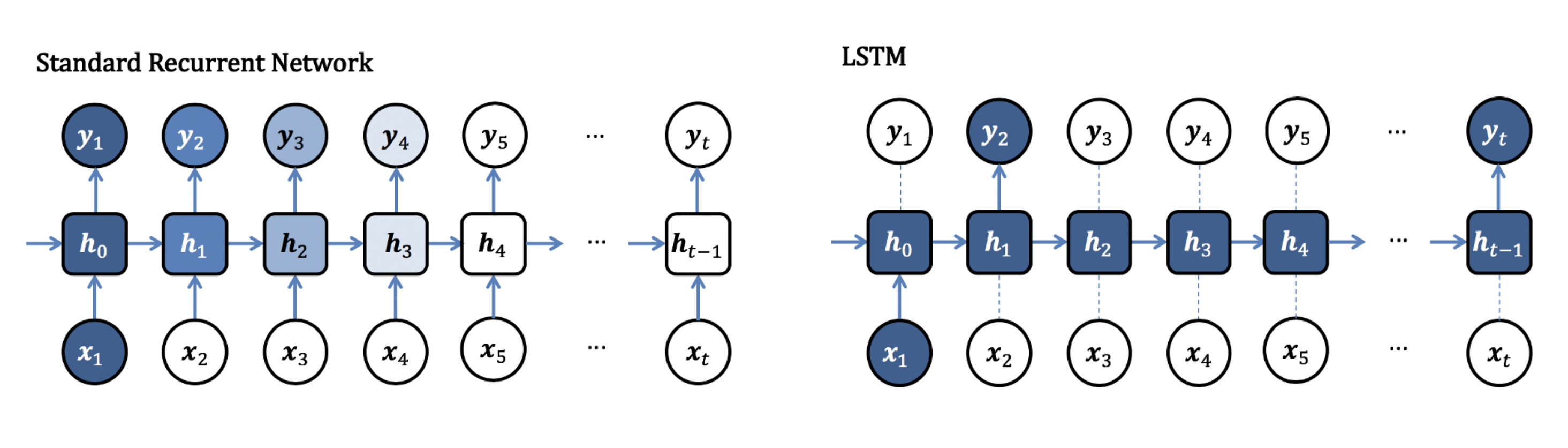}
    \caption{The vanishing gradient problem and loss of context. Demonstration of preservation of contextual information in an LSTM network \citep{graves}}
    \label{o:1}
\end{figure}

Recurrent neural networks are neural networks designed for sequence processing. They process one element of a sequence at a time while storing the information about its relationships to the previous element(s) of the sequence. Although their main goal is to learn long-term relationships, theoretical and empirical evidence has shown that it is difficult for them to learn to store information for a very long time, leading to a problem called the vanishing gradient problem. Therefore, a special type of recurrent neural network has become popular -- namely, Long Short-Term Memory (LSTM) \citep{hochreiter1997long} -- which can reflectively solve the vanishing gradient problem (see Figure \ref{o:1}) \citep{graves}.

An LSTM remembers inputs for a long time through the use of hidden units called memory cells. These cells function as gated leaky neurons, which associate with each other in the following step. In practice, this means that it adds an external signal to its real state; this connection is encoded multiple times by another neuron, which decides when to delete the content from its memory.

A Bidirectional Long Short-Term Memory network (BiLSTM) is a specific type of LSTM network.  BiLSTMs consist of two individual hidden layers. The first layer processes the input sequence in a forward direction, while the second hidden layer processes the sequence in a backward direction. The hidden layers then merge their outputs in the output layer. Consequently, the output layer can access both the past and the future context of the current event in the sequence. LSTM and its bidirectional variants have been proved to be very useful. They can learn how and when they can forget certain information, and they can learn not to use some gateways in their architecture. Faster learning rates and better performance are the advantages of BiLSTM networks \citep{schuster1997bidirectional, lstm_kamilya}.

\subsubsection{Gated Recurrent Unit}
In 2014,  a new type of network similar to LSTM was introduced. Its task is to reduce the number of parameters that the LSTM block contains. This new type, known as a Gated Recurrent Unit (GRU) block, contains two gates: $r$ (reset gate) and $z$ (update gate). The individual parts of the GRU block perform the following functions:
\begin{itemize}
    \item The current activation, $h$, is a linear interpolation between the previous activation and the candidate activation;
    
    \item the candidate activation, $\hat{h}$, is a candidate for further activation of $h$;
    
    \item the reset gate, $r$, performs a similar role as the forget gate in the LSTM block; and
    
    \item the update gate, $z$, determines how much the unit changes its activation.
    
\end{itemize}

\begin{figure}[ht!]
    \centering
    \includegraphics[width=0.95\textwidth]{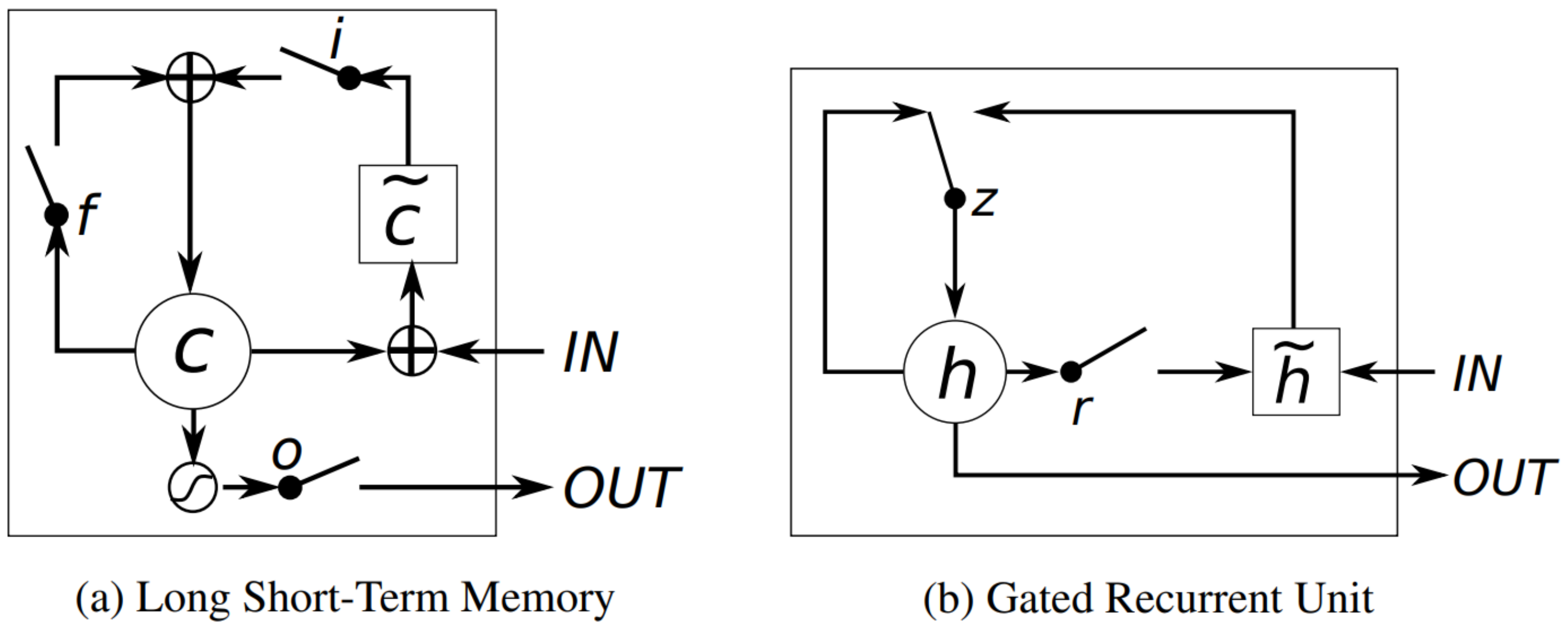}
    \caption{Difference between LSTM and GRU. In sub-image (a): $i$, $f$, and $o$ represent input, forget, and output gates, respectively, while $c$ and $\hat{c}$ represent the current state of the memory block and the new state of the memory block, respectively. In sub-image (b): $r$ and $z$ represent the reset and update gates, while $h$ and $\hat{h}$ are the activation and candidate activation, respectively \citep{Chung2014EmpiricalEO}.}
    \label{o:2}
\end{figure}

GRUs do not contain memory cells and the number of parameters that each block contains is significantly lower than that in LSTM (see Figure \ref{o:2}).

\subsection{Modeling and evaluation metrics}
In the modeling phase, we created various neural networks, from the simplest feed-forward to a bidirectional LSTM in combination with CNN. The structures of the most successful models are described below (see subsection \ref{experiments}). We used the following functions, hyper-parameters, and evaluation metrics for all the models created:

\paragraph{Activation functions} 

We used the ReLU activation function in the hidden layers of the neural networks. The output of the ReLU function can be represented as:

\begin{equation}\label{relu_vzorec}
    	   f(x) = \max(0,x).
\end{equation}

We used the softmax function in the output layer. For classification into two classes, the use of softmax is equivalent to the use of a sigmoidal activation function and provides a categorical probability distribution. Mathematically, the softmax function is represented as:

\begin{equation}\label{softmax_vzorec}
    	   f(x_{k}) = \frac{e^{z_{i}}}{\sum ^{K}_{j=1}e^{z_{j}}}.
\end{equation}

\paragraph{Loss function}
The loss function estimates the error the model makes during training, which is important for the learning process. The backpropagation method updates the weights of the neurons in a given layer in such a way that errors (i..e, losses) are decreased during the next evaluation. In our case, we used categorical cross-entropy \citep{CE}, which calculates the entropy-based difference between two probability distributions. The value of categorical cross-entropy increases if the assumed probability differs from the actual class (category). Generally, for $M$ classes, we can calculate the categorical CE as follows: 
\begin{equation}\label{referencia_0}
\mathrm{CE} = -\sum_{i}^{M} y_{i}\log (\widehat{y}_{i}),
\end{equation}
where $y$ is the actual value, $ \hat{y} $ is the predicted value, and $M$ is the number of classes.
    
\paragraph{Optimization}
For the optimization procedure, which searches for optimal parameters for the reduction of the loss rate, we used the Adam optimizer \citep{kingma2014adam}. Adam is an adaptive learning rate optimization algorithm designed for neural network training. The name Adam is derived from the term adaptive moment estimation. It uses a quadratic gradient to change the learning rate, as well as the momentum based on the moving average of the gradient. Adam is an adaptive learning method, which means that it calculates an individual learning rate for each parameter. The parameters $\omega^{(t)}$ and the loss function $L^{(t)}$ are given below, where $ t $ indexes the current iteration of learning. With this method, the parameters are updated as follows:

\begin{equation}\label{referencia_1}
m_{\omega }^{(t+1)} \leftarrow  \beta _{1}m_{\omega }^{(t)} + (1 - \beta _{1})\bigtriangledown_{\omega }L^{(t)},
\end{equation}

\begin{equation}\label{referencia_2}
v_{\omega }^{(t+1)} \leftarrow  \beta _{2}v_{\omega }^{(t)} + (1 - \beta _{2})(\bigtriangledown_{\omega }L^{(t)})^{2},
\end{equation}
\begin{equation}\label{referencia_3}
\widehat{m}_{\omega} = \frac{m_{\omega }^{(t+1)}}{1 - \beta _{1}^{(t+1)}},
\end{equation}
\begin{equation}\label{referencia_4}
\widehat{v}_{\omega} = \frac{v_{\omega }^{(t+1)}}{1 - \beta _{2}^{(t+1)}},
\end{equation}
\begin{equation}\label{referencia_5}
\omega ^{(t+1)} \leftarrow \omega^{(t)} - \eta \frac{\widehat{m}_{w}}{\sqrt{\widehat{v_{\omega }}}+\epsilon},
\end{equation}
where $ \epsilon $ is a small scalar that is used to prevent division by zero, and $ \beta_1 $ and $ \beta_2 $ are damping factors. The default settings are $ \beta_1 = 0.9$, $ \beta_2 = 0.999$, and $\epsilon = 10^{- 8}$, with a learning speed of 0.001.

\paragraph{Regularization} 
For regularization during training, we used Dropout \citep{srivastava2014dropout}. With each iteration, Dropout randomly selects nodes and deletes them along with all the inputs and outputs that belong to them. Each iteration contains different nodes, which results in different outputs. We can look at this as an ensemble method combining a set of models, which usually performs better than the simple model obtained without dropout. The probability of selecting the number of nodes to be omitted is the regularization dropout parameter.

We also used another technique known as Early stopping, in which some of the data from the data set intended for training is used as a validation set. This is a method that monitors whether the results on the validation set deteriorate. If so, model training is stopped.

\paragraph{Metrics} To evaluate the models, we used standard metrics for classification (i.e., accuracy, precision, recall, and F1 score). Such metrics are easy and straightforward to obtain for a binary classification problem from the confusion matrix, which contains counts of examples for four different resulting evaluation cases. For example, let us assume that our class for evaluation is a detached binary system. Then, TP (True Positive) examples are those that we know are detached and which the model predicts are detached. TN (True Negative) examples are predicted as not detached and they are not detached. FN (False Negative) examples are predicted not to be detached but are detached. FP (False Positive) examples are predicted to be detached but are not detached. The metrics used in the evaluation are defined as follows:
\begin{itemize}
 \item   Accuracy = (TP+TN) / (TP+FP+FN+TN),
 \item   Precision = TP / (TP+FP),
 \item   Recall = TP / (TP+FN),
 \item  F1 score = 2 $\ast$ (Precision $\ast$ Recall) / (Precision + Recall).
\end{itemize}

The results reported in the evaluation are micro-averaged. Micro-averaging first computes the confusion matrix (see Figure \ref{o:matrix}) for all classes and then computes the overall metrics. Micro-averaging may be preferred in the case of imbalance classes, which is the case in this study, as the particular classes have different numbers of evaluated examples. Micro-averaged precision, recall, and F1 score are computed as follows (with indexing of $|C|$ classes):

\begin{equation}
\mathrm{Precision_{micro}}= \frac{\sum_{i=1}^{|C|} \mathrm{TP}_{i}} {\sum_{i=1}^{|C|} (\mathrm{TP}_{i} + \mathrm{FP}_{i})},
\end{equation}
\begin{equation}
\mathrm{Recall_{micro}}= \frac{\sum_{i=1}^{|C|} \mathrm{TP}_{i}} {\sum_{i=1}^{|C|} (\mathrm{TP}_{i} + \mathrm{FN}_{i})},
\end{equation}
\begin{equation}
\mathrm{F1\ score_{micro}}= 2\ast\frac{\mathrm{Precision_{micro}} \ast \mathrm{Recall_{micro}}} {\mathrm{Precision_{micro}} + \mathrm{Recall_{micro}}}.
\end{equation}

\begin{figure}[!h]
    \begin{center}
    {
        \offinterlineskip
        \raisebox{-2.7cm}[0pt][0pt]{
            \parbox[c][2.5pt][c]{4.5cm}{\hspace{-3.1cm}{\textbf{Actual}}\\[10pt]}}\par

        \hspace*{0.5cm}\MyHBox[\dimexpr3.4cm+6\fboxsep\relax]{Predicted}\par

        \hspace*{0cm}\MyHBox{1}\hspace*{2cm}\MyHBox{0}\par

        \MyTBox{1}{True Positive}{False Negative}

        \MyTBox{0}{False Positive}{True Negative}

    }
\end{center}
\caption{Confusion matrix}
\label{o:matrix}
\end{figure}

\section{Experiments and results}\label{experiments}

Using the ELISa software, we obtained enough data for training and validation. During training, the model achieved 100\% accuracy on the validation set in each model. However, overfitting proved to be an issue for the first experiments. While the accuracy rate on the validation set was 100\%, the resulting values were different on the evaluation set (i.e., on the observation light curves of binaries). We needed to carefully set up more experiments before we found the right balance of architectural and hyper-parameter settings without overfitting issues. The resulting settings of the hyper-parameters of the two most successful models are given in Table \ref{tab:hyperparam}.

\begin{table}[!h]
	\centering
	\caption{Hyper-parameters used to train our best classifier.}
	\label{tab:hyperparam}
	\begin{tabular}{ccccc} 
		\toprule
		 \textbf{Hyper-parameter} & \textbf{Values}\\
		\midrule
	     Epochs & 5\\
         Batch Size & 32\\
		 Learning rate & $0.001$\\
		 Dropout rate &  20\% \\
		\bottomrule
	\end{tabular}
\end{table}
In every model, the last output layer consisted of two neurons fully connected to the previous part of the architecture. The output layer has a softmax activation function and provides an output for two classes (detached and over-contact). In the following, we describe our architectures and how they differed before the mentioned output layer.   
For the first model, we tested a simple CNN consisting of two convolutional blocks, a fully connected layer with 64 neurons, and regularization dropout of 25\%. The first convolution block contained a 1D convolutional layer with 64 filters, kernel size 20, and a max-pooling layer with pool size 2. The second convolution block contained a 1D convolution layer with 32 filters and kernel size 10. The accuracy of this model reached 81\%.
The next two models were simple LSTM and GRU networks. The LSTM model contained an LSTM layer with 64 units and three fully connected layers with 64, 64, and 32 neurons. We chose the same sequence in the GRU model, where the GRU layer contained 64 units followed by three fully connected layers with the same number of neurons. The GRU model achieved ten percent better accuracy (94\%) for the best model than the LSTM (84\%).

Our next experiments focused on block-based architectures using a combination of LSTM and convolutional layers. In the first case, we tried a model where the output of the convolution block was used as input for the LSTM layer (Model 4). This model included a 1D convolutional layer with 32 filters with kernel size 3, a max-pooling layer with pool size 2, and dropout regularization of 20\%.  The output of the convolution part provided an input for the LSTM layer, with 64 units and a fully connected layer with 32 neurons. Model 4 achieved better results than either of the CNN or LSTM. Table \ref{tab:sum} provides a summary of the Model 4 architecture. This model achieved 96\% accuracy. 

\begin{table}[!h]
\centering
\caption{Summary of Model 4 architecture.}
\begin{tabular}{lcr}
\toprule                                    
\textbf{Layer (type)  }                                & \textbf{Output Shape }  & \textbf{Parameters}                                             \\ \midrule
InputLayer                       & (None, 100, 1) & 0           \\
Conv1D                           & (None, 98, 32) & 128         \\
MaxPooling1  & (None, 49, 32)          &      0    \\
Dropout                          & (None, 49, 32) & 0           \\
LSTM                                & (None, 49, 64) & 24,832    \\
Flatten                          & (None, 3136)   & 0           \\
Dense                              & (None, 32)     & 100,384    \\ 
Dense                              & (None, 2)     & 66    \\ \midrule
\textbf{Total params:     }                   &   125,410               &          \\
\textbf{Trainable params: }                    &    125,410             &          \\
\textbf{Non-trainable params: }                     &     0          &         \\ \bottomrule
\end{tabular}
\label{tab:sum}
\end{table}

\begin{figure}[!h]
    \centering
    \includegraphics[width=\textwidth]{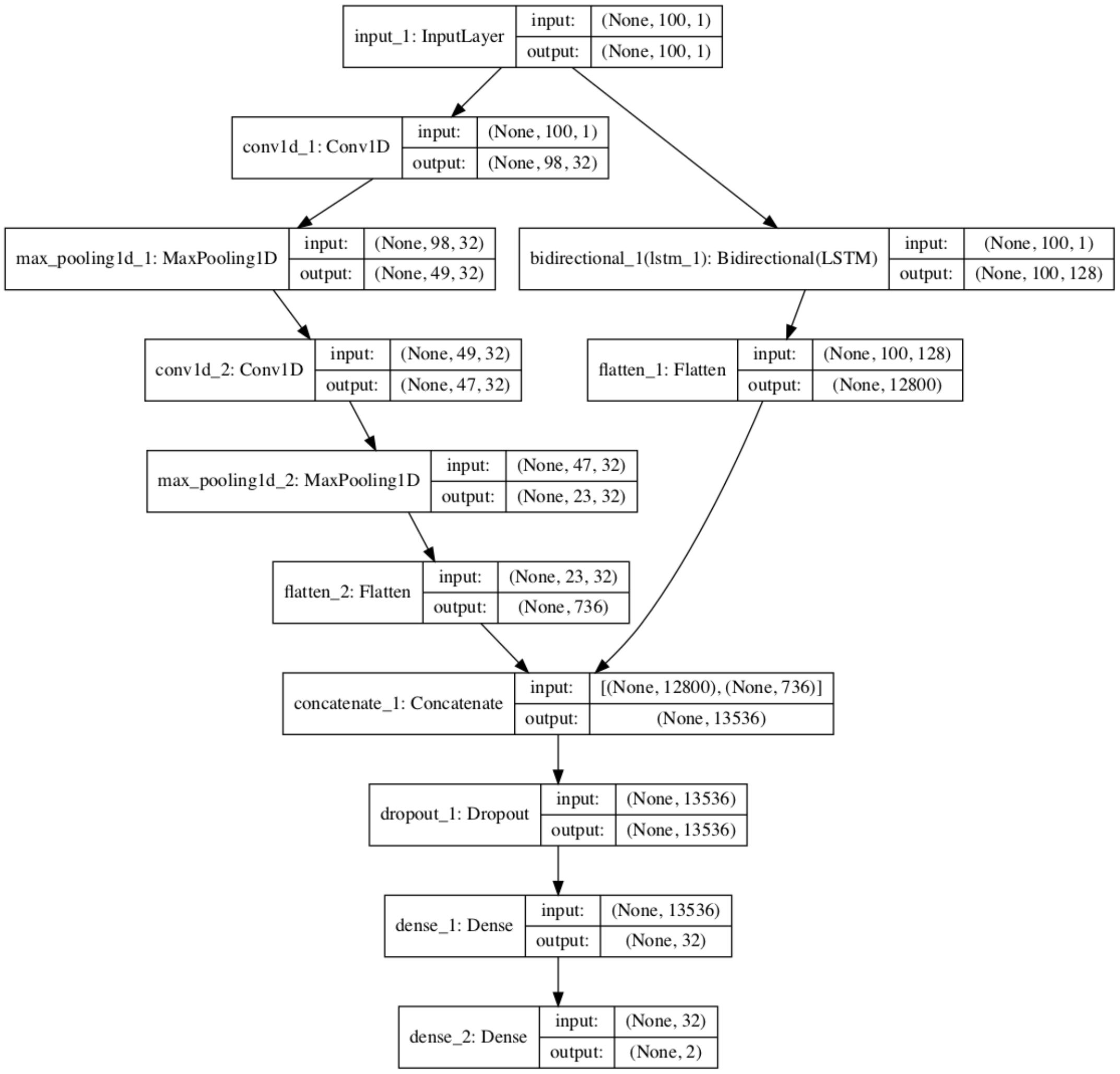}
    \caption{Architecture of Model 6 -- combination of bidirectional LSTM and convolutional blocks.}
    \label{o:model}
\end{figure}

\begin{figure}[!h]
    \centering
    \includegraphics[width=.47\textwidth]{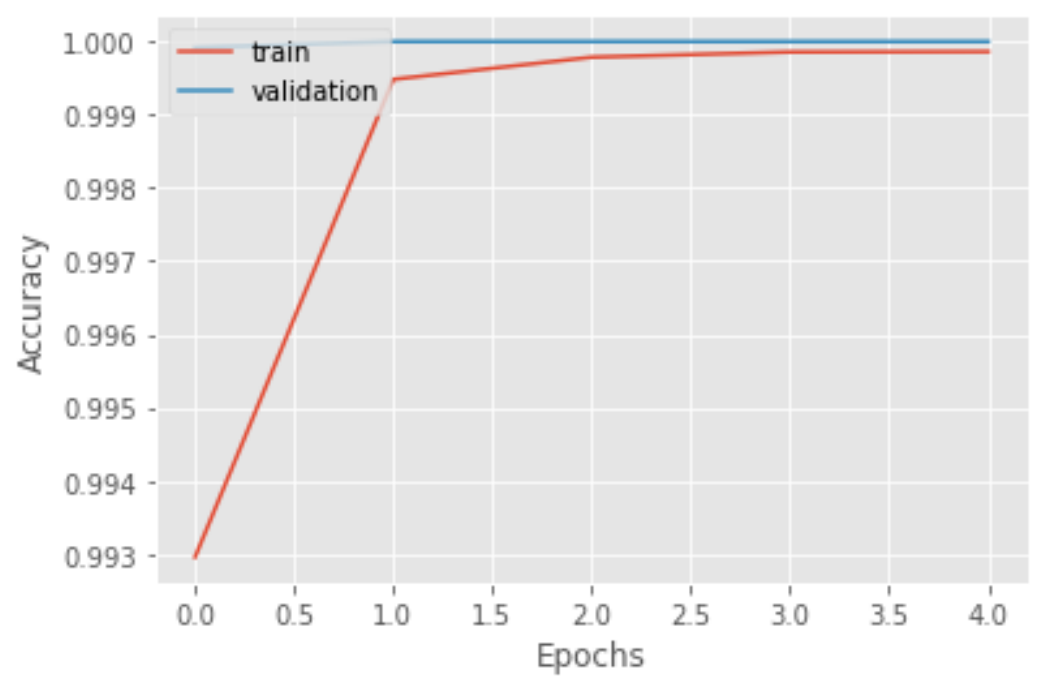}
    \includegraphics[width=.47\textwidth]{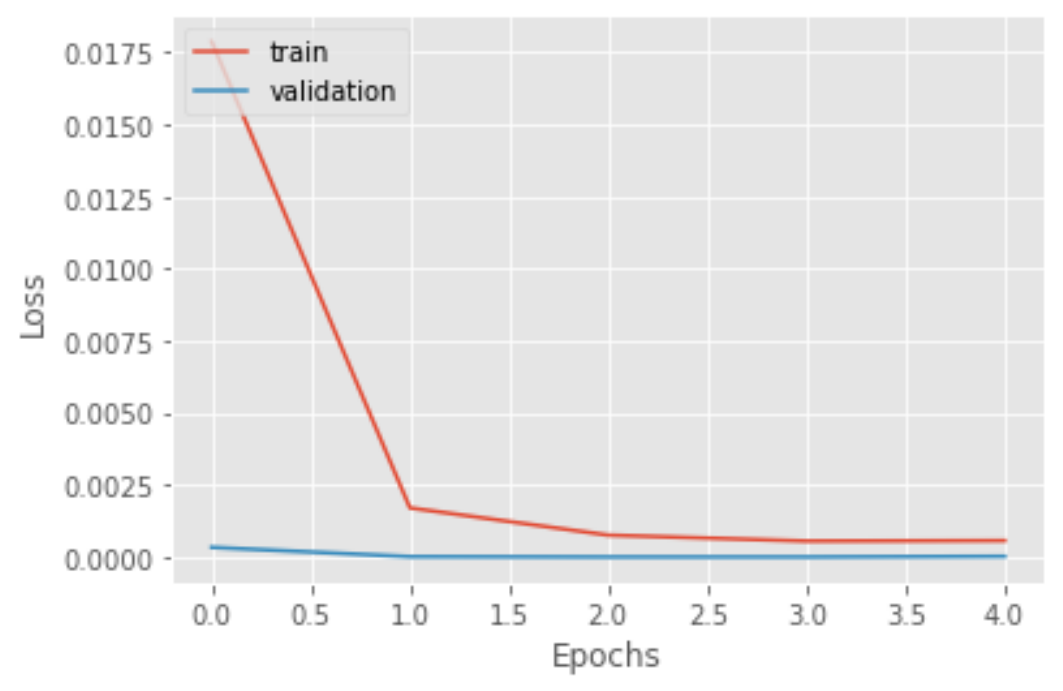}
    \caption{Learning curves showing the accuracy and loss for Model 6 in individual epochs for the sequence of training phases.}
    \label{fig:train_curves}
\end{figure}

Then, we tried another approach that processed the input using parallel convolutional and LSTM blocks.  The outputs of these were then combined using a Concatenation layer (Model 5 and Model 6).
For Model 5, the first block consisted of an LSTM layer with 64 units, while the second block was a convolution layer with 64 filters and kernel size 3. The outputs of the two blocks were combined to provide the input for a fully connected layer with 64 hidden neurons.  We also used 20\% dropout for this architecture. This model showed promising results but was unable to capture some of the detailed aspects of the light curves. 
Therefore, for the last model (Model 6), we decided to use a similar architecture to the previous one. However, for the first parallel part, we used a bidirectional LSTM layer while, for the second part, we used two convolutional sub-blocks. We expected better feature selection on the light curve from the CNN part and also that the biLSTM could better capture sequence-related aspects due to its bidirectional processing. With this model, we were able to achieve 98\% micro-precision, recall, and F1 score. Figure \ref{o:model} shows the overall structure of this model. Figure \ref{fig:train_curves} shows learning curves of Model 6 in individual epochs on the training set and their respective accuracy and loss.

The average values for the evaluation metrics of all models are displayed in Table \ref{tab:result}. For these results, all 100 observational light curves were classified as either detached or over-contact, with semi-detached cases (14 examples) evaluated as detached. 

\begin{table}[!h]
\caption{Results of individual models on observational light curves of eclipsing binary stars.}
\centering
\begin{tabular}{lcccc}
\toprule
                                                                      \textbf{Model}       & \textbf{Precision}     & \textbf{Recall}        & \textbf{F1 score }         \\ \midrule
\begin{tabular}[c]{@{}l@{}}Model 1  (CNN)\end{tabular}               & 0.81          & 0.81         & 0.81          &           \\            

\begin{tabular}[c]{@{}l@{}}Model 2 (LSTM)\end{tabular}               & 0.84          & 0.84          & 0.84          &          \\

\begin{tabular}[c]{@{}l@{}}Model 3 (GRU)\end{tabular}               & 0.94        & 0.94        & 0.94          &        \\

\begin{tabular}[c]{@{}l@{}}\textbf{Model 4  (CNN, LSTM)}\end{tabular}               & \textbf{0.96 }         & \textbf{0.96}          & \textbf{0.96}          &          \\

\begin{tabular}[c]{@{}l@{}}Model 5 (CNN + LSTM)\end{tabular}            & 0.95          & 0.95          & 0.95         &        \\

\textbf{\begin{tabular}[c]{@{}l@{}}Model 6 (biLSTM + 2x CNN) \end{tabular}} & \textbf{0.98} & \textbf{0.98} & \textbf{0.98}

\\ \bottomrule
\end{tabular}
\label{tab:result}
\end{table}

Next, we compared  the two best classifiers in detail. Model 4 classified four light curves incorrectly: Three of them were incorrectly predicted as over-contact eclipsing binary stars and one as detached. Compared to the second-most successful model, Model 6 better classified both classes. The results for the individual classes using the best model (Model 6) are displayed in Table~\ref{tab:result6}. From the confusion matrix, it is clear that two of the 100 curves were classified incorrectly. In particular, these two curves were wrongly predicted as over-contact.

\begin{figure}[!ht]
    \centering
    \includegraphics[width=0.8\textwidth]{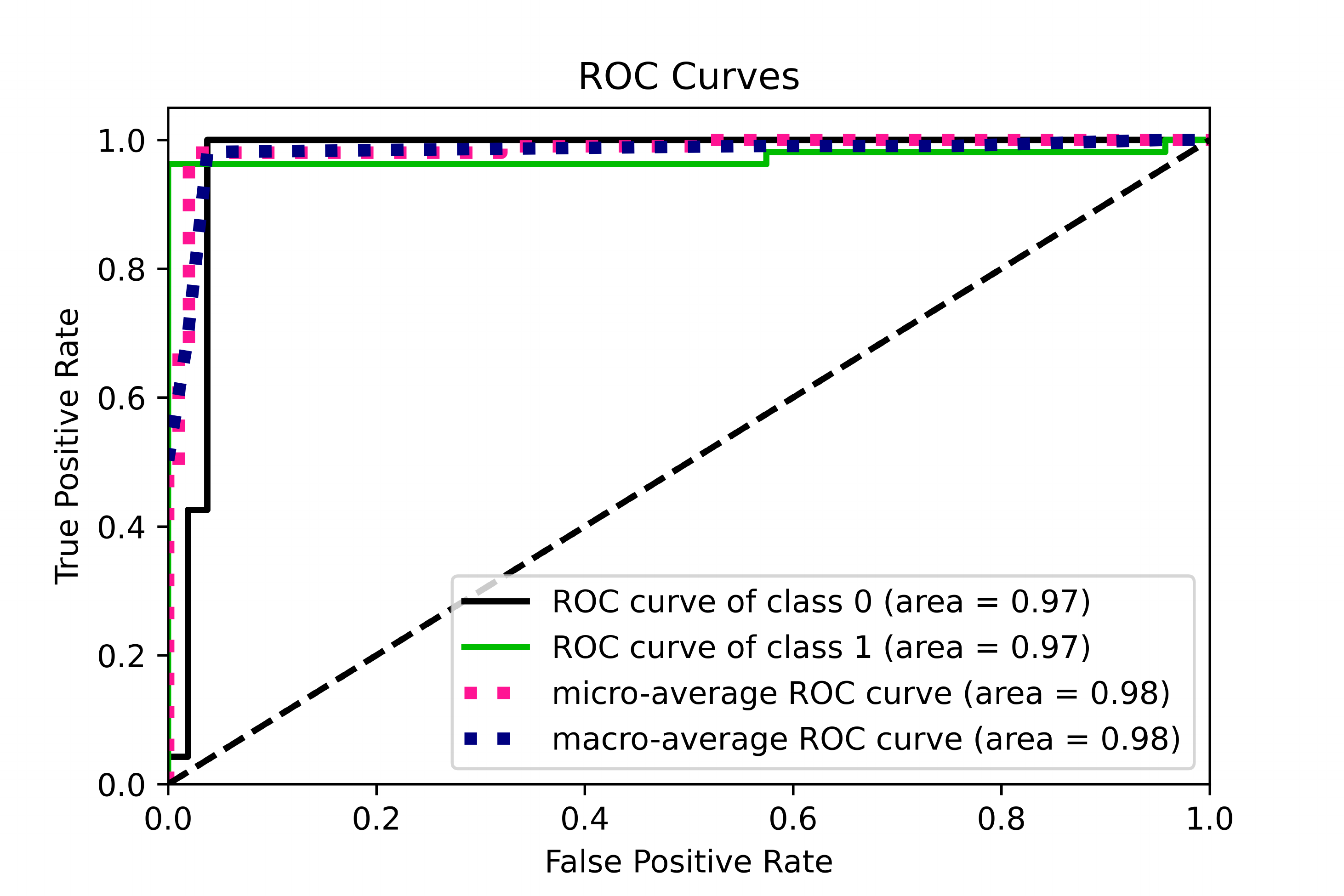}
    \caption{ROC curves that show the performance of classification for Model 6 at all classification thresholds.}
    \label{fig:roc}
\end{figure}

The confusion matrices and the results for the two best classifiers (Models 4 and 6) are given in Table~\ref{tab:result6} (for Model 4) and Table~\ref{tab:result3} (for Model 6). Figure \ref{fig:roc} shows the receiver operating characteristic curves (ROC curves) for Model 6 for the individual classes and the micro and macro average ROC curves for this model. Figure~\ref{o:result6} shows the incorrectly predicted light curves for Model 6. Similarly, Figure~\ref{o:result3} provides these for Model 4.

\begin{table}[!h]
\centering
\caption{Results for each class and confusion matrix -- Model 4}
\begin{tabular}{lcccc}
\toprule
      \textbf{Type}      & \textbf{Precision} & \textbf{Recall} & \textbf{F1 score} & \textbf{Support} \\ \midrule
      \textbf{Over-contact} & 0.94      & 0.98   & 0.96     & 47      \\ 
\textbf{Detached}       & 0.98      & 0.94   & 0.96     & 53      \\
\bottomrule
\end{tabular}

\vspace{0.4cm}

\centering
\begin{tabular}{@{}cccc@{}}
\toprule
\multicolumn{1}{l}{}                                  & \multicolumn{3}{c}{\textbf{Predicted}} \\ \midrule
\multicolumn{1}{c|}{\multirow{3}{*}{\textbf{Actual}}} &              & \textbf{Over-contact}     & \textbf{Detached}     \\
\multicolumn{1}{c|}{}                                 & \textbf{Over-contact }   &        46      &     1    \\
\multicolumn{1}{c|}{}                                 & \textbf{Detached}         &    3          &         50 \\ \bottomrule
\end{tabular}
\label{tab:result3}
\end{table}

\begin{table}[!h]
\centering
\caption{Results for each class and confusion matrix -- Model 6}
\begin{tabular}{lcccc}
\toprule
      \textbf{Type}      & \textbf{Precision} & \textbf{Recall} & \textbf{F1 score} & \textbf{Support} \\ \midrule
      \textbf{Over-contact} & 0.96      & 1.00   & 0.98     & 47      \\ 
\textbf{Detached}       & 1.00      & 0.96   & 0.98     & 53      \\
\bottomrule
\end{tabular}

\vspace{0.4cm}

\centering
\begin{tabular}{@{}cccc@{}}
\toprule
\multicolumn{1}{l}{}                                  & \multicolumn{3}{c}{\textbf{Predicted}} \\ \midrule
\multicolumn{1}{c|}{\multirow{3}{*}{\textbf{Actual}}} &              & \textbf{Over-contact}     & \textbf{Detached}     \\
\multicolumn{1}{c|}{}                                 & \textbf{Over-contact }   &        47      &     0     \\
\multicolumn{1}{c|}{}                                 & \textbf{Detached}         &    2          &         51 \\ \bottomrule
\end{tabular}
\label{tab:result6}
\end{table}

We implemented all models in Python, using the Tensorflow \citep{abadi2016tensorflow} and Keras \citep{keras} packages for neural network architectures. The codes are available online on GitHub\footnote{https://github.com/VieraMaslej/eclipsing-binary-stars}. The experiments were conducted on a PC equipped with a 4-core Intel Xeon processor clocked at 4 GHz and an NVIDIA Tesla K40c GPU with 12 GB memory.

\section{Analysis and discussion of results}
 
\begin{figure}[ht!]
    \centering
    \includegraphics[width=0.49\textwidth]{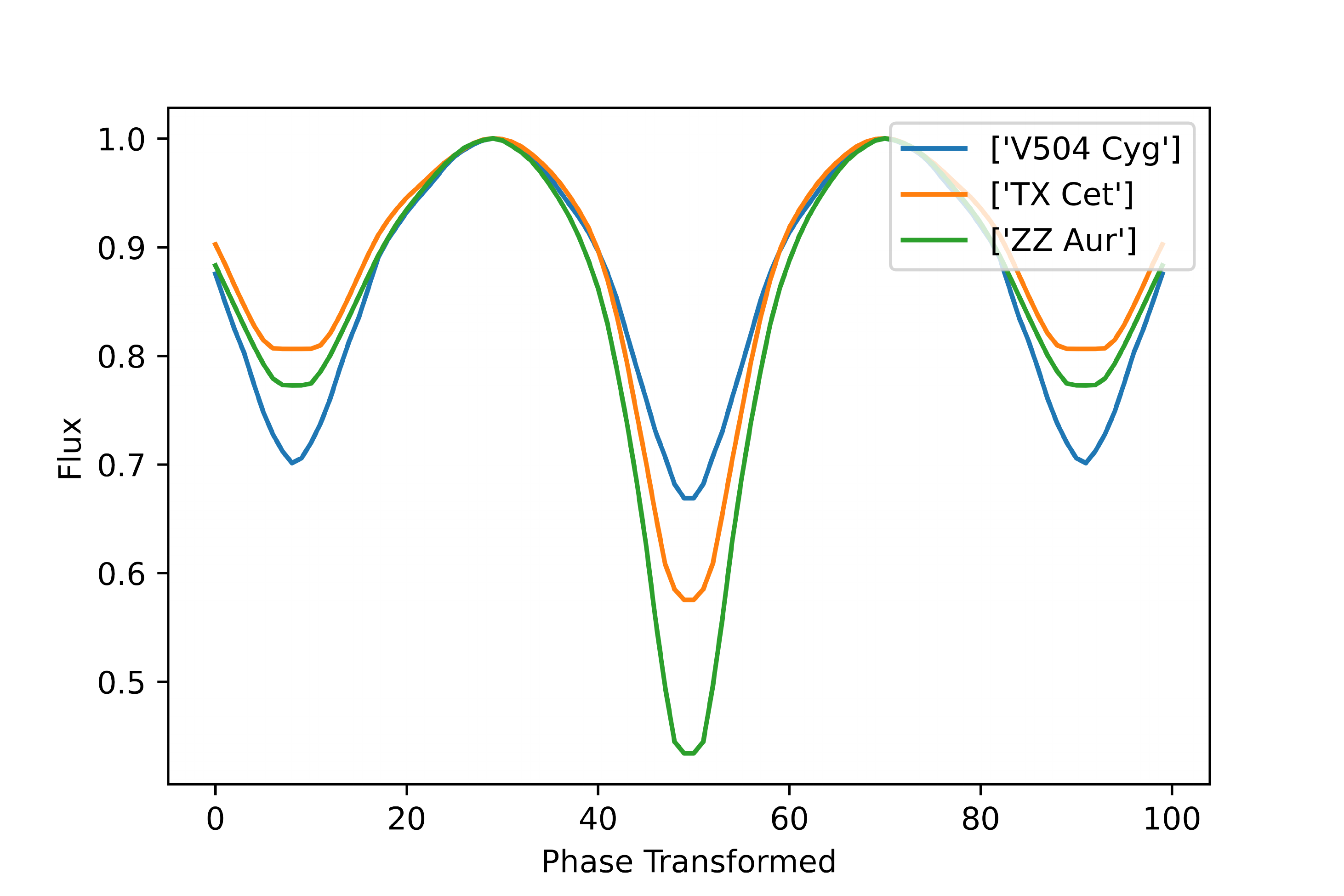}
    \includegraphics[width=0.49\textwidth]{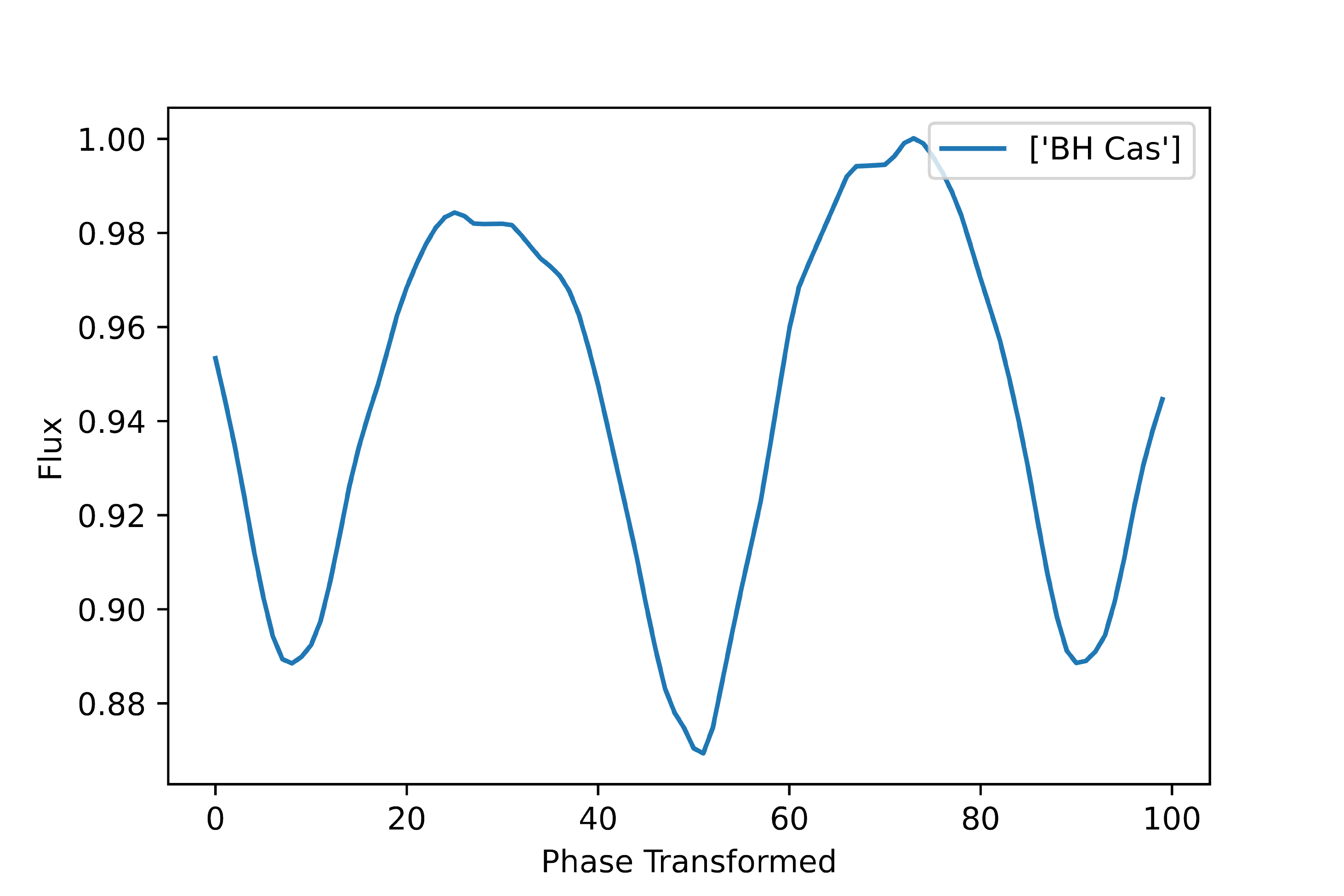}
    \caption{Misclassified light curves of binary stars -- Model 4. The first graph shows the curves predicted as over-contact but, in fact, they were in the detached class (semi-detached). The second graph shows one curve incorrectly predicted as a detached eclipsing binary star.}
    \label{o:result3}
\end{figure}

We analyzed the results of the best architectures in more detail. Model 4 (the serialized combination of CNN and LSTM) classified the light curves of V504 Cyg, TX Cet, and ZZ Aur incorrectly into the over-contact class. These three cases are all semi-detached binaries, which were considered as detached. The model also incorrectly predicted one light curve as detached; namely, BH~Cas (see Figure \ref{o:result3}).

\begin{figure}[ht!]
\centering
   \includegraphics[width=0.55\textwidth]{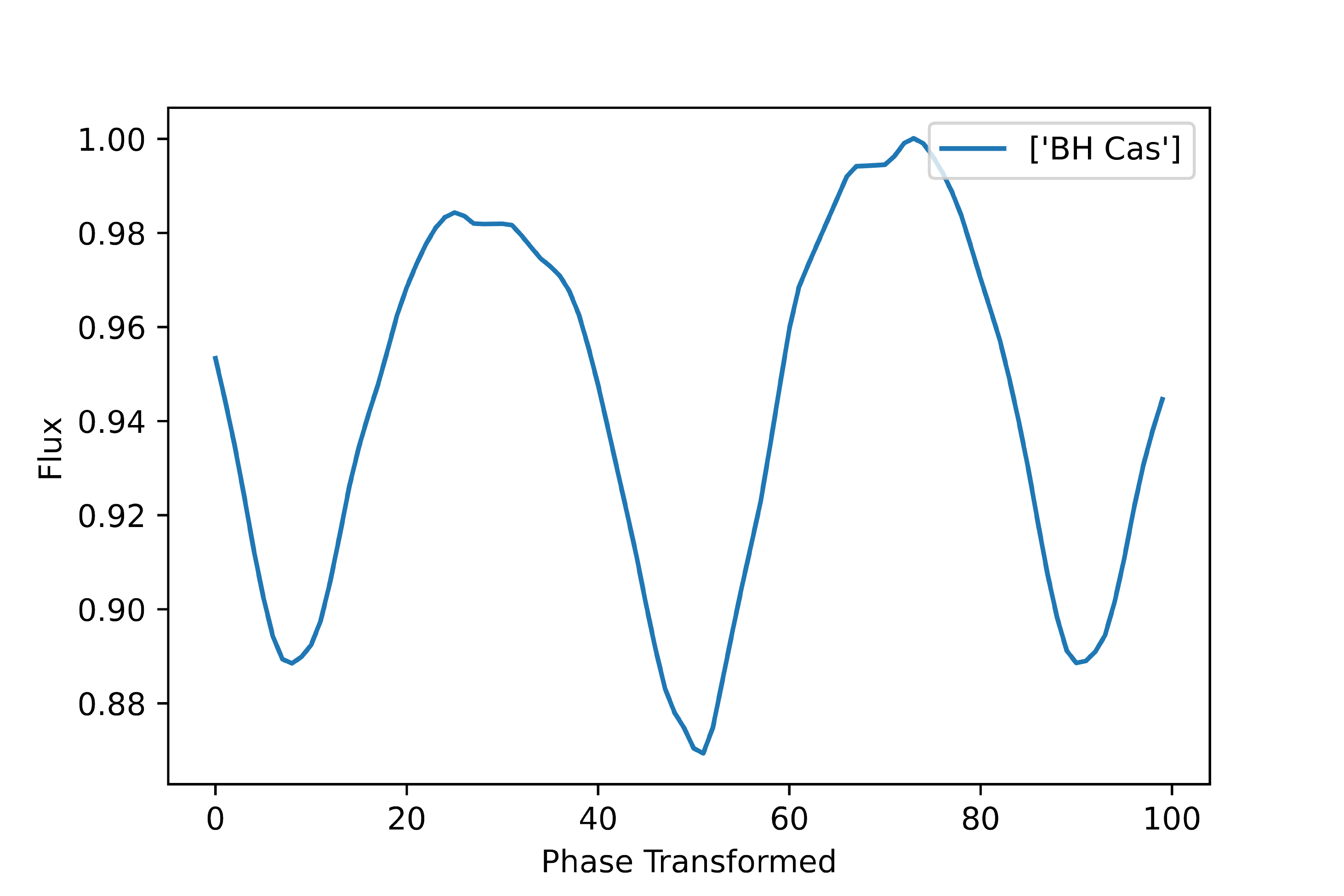}

    \caption{Misclassified light curve -- Model 4 evaluated without observed semi-detached light curves of binaries. The graph shows the light curve predicted as the detached class while, in fact, it was the over-contact class.}
  \label{o:resulttop}
\end{figure}

If we did not take into account the observations of semi-detached light curves, the accuracy of Model 4 would be 99\% (see Table \ref{tab:resulttop}). In this case, only one light curve -- BH Cas -- was incorrectly predicted (see Figure \ref{o:resulttop}).

\begin{table}[!h]
\centering
\caption{Results for each class and confusion matrix -- Model 4 evaluated without observed semi-detached light curves of eclipsing binaries.}
\begin{tabular}{lcccc}
\toprule
      \textbf{Type}      & \textbf{Precision} & \textbf{Recall} & \textbf{F1 score} & \textbf{Support} \\ \midrule
      \textbf{Over-contact} & 1.00      & 0.98   & 0.99     & 47      \\ 
\textbf{Detached}       & 0.97      & 1.00   & 0.99     & 39     \\
\bottomrule

\end{tabular}

\vspace{0.4cm}

\centering
\begin{tabular}{@{}cccc@{}}
\toprule
\multicolumn{1}{l}{}                                  & \multicolumn{3}{c}{\textbf{Predicted}} \\ \midrule
\multicolumn{1}{c|}{\multirow{3}{*}{\textbf{Actual}}} &              & \textbf{Over-contact}     & \textbf{Detached}     \\
\multicolumn{1}{c|}{}                                 & \textbf{Over-contact }   &        46      &     1    \\
\multicolumn{1}{c|}{}                                 & \textbf{Detached}         &    0         &         39 \\ \bottomrule
\end{tabular}
\label{tab:resulttop}
\end{table}

\begin{figure}[ht!]
    \centering
    \includegraphics[width=0.55\textwidth]{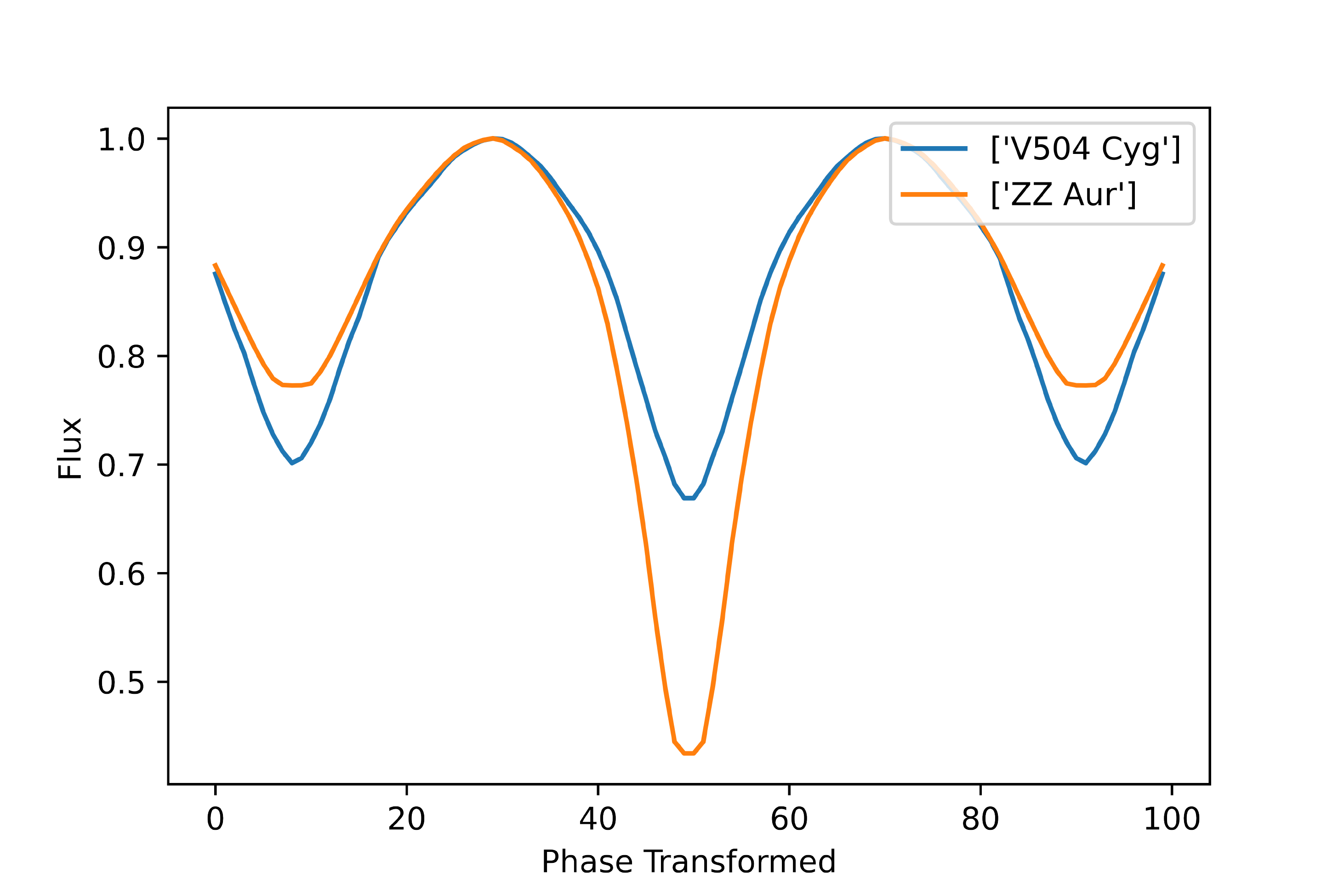}
    \caption{Misclassified light curves of binary stars -- Model 6. The graph shows the curves predicted as over-contact class which, in fact, were in the detached class (semi-detached).}
    \label{o:result6}
\end{figure}

If we focus on the misclassified light curves in Figure \ref{o:result6}, we can see that Model 6 predicted V 504 Cyg and ZZ Aur as over-contact. However, both were semi-detached binaries (evaluated as detached). If we did not take into account the semi-detached examples, Model 6 (biLSTM and CNN) classified all 86 curves correctly. The results for each class and the confusion matrix for Model 6 evaluated without observed semi-detached light curves of eclipsing binaries are shown in Table \ref{tab:bilstmbest}.

\begin{table}[!h]
\centering
\caption{Results for each class and confusion matrix -- Model 6 evaluated without observed semi-detached light curves of eclipsing binaries.}
\begin{tabular}{lcccc}
\toprule
      \textbf{Type}      & \textbf{Precision} & \textbf{Recall} & \textbf{F1 score} & \textbf{Support} \\ \midrule
      \textbf{Over-contact} & 1.00      & 1.00   & 1.00     & 47      \\ 
\textbf{Detached}       & 1.00      & 1.00   & 1.00     & 39     \\
\bottomrule

\end{tabular}

\vspace{0.4cm}

\centering
\begin{tabular}{@{}cccc@{}}
\toprule
\multicolumn{1}{l}{}                                  & \multicolumn{3}{c}{\textbf{Predicted}} \\ \midrule
\multicolumn{1}{c|}{\multirow{3}{*}{\textbf{Actual}}} &              & \textbf{Over-contact}     & \textbf{Detached}     \\
\multicolumn{1}{c|}{}                                 & \textbf{Over-contact }   &        47      &     0    \\
\multicolumn{1}{c|}{}                                 & \textbf{Detached}         &    0         &         39 \\ \bottomrule
\end{tabular}
\label{tab:bilstmbest}
\end{table}

\section{Conclusions and future work}
By using the synthetic data generator ELISa, we obtained a sufficient amount of synthetic light curves for eclipsing binary stars, in order to train models based on several neural network architectures. We further collected 100 observational light curves of eclipsing binary stars from published studies and used them to evaluate the created models. Using deep learning methods, we created six models and compared and analyzed their results.

Based on light curve data, our best-performing classifier could automatically classify observed binary stars into two classes (detached and over-contact) with 100\% accuracy. When we included the light curves of semi-detached binaries into the detached class, our model -- based on a parallel combination of bidirectional Long Short-Term Memory and convolutional layers -- could predict the correct class with 98\% accuracy.

For future work, it would be appropriate to generate curves for three classes (detached, semi-detached, over-contact) and then train a classifier that also recognizes the third class; that is, semi-detached. Another extension could be based on theoretical models of the occurrence of spots on binary stars, in order to create a complex classifier that is able to classify stars according to both the morphology and influence of spots. For these purposes, however, it would be necessary to extend the synthetic data generator and also to expand the set and the properties of the observational data.

\section*{CRediT authorship contribution statement}
\textbf{M. \v{C}okina:} Conceptualization, Writing - review \& editing, Software, Validation. \textbf{V. Maslej-Kre\v{s}\v{n}\'akov\'a:} Methodology, Software, Writing - Original Draft, Visualization. \textbf{P. Butka:} Writing - review \& editing, Validation, Supervision. \textbf{\v{S}. Parimucha:} Writing - review \& editing, Supervision.

\section*{Declaration of competing interest}
The authors declare that they have no known competing financial interests or personal relationships that could have appeared to influence the work reported in this paper.

\section*{Acknowledgements}
This  research  was  supported  by  the  Slovak  Research  and  Development Agency under the contracts No. APVV-15-0458 and No. APVV-16-0213, and also by the Slovak VEGA grant No. 1/0685/21.

\bibliography{mybibfile}
\end{document}